\documentclass[reqno,a4paper,final,12pt]{amsart}
\usepackage[british]{babel}
\usepackage{amssymb,amstext,amsthm,eucal}
\usepackage[margin=1in]{geometry}
\usepackage{graphicx}
\usepackage{tikz}
\usepackage{array}
\usepackage{slashed}
\usepackage[utf8]{inputenc}
\usepackage[notref,notcite]{showkeys}
\usepackage[small]{eulervm}
\usepackage{tgpagella}
\usepackage{multicol}
\usepackage{float}
\usepackage{enumerate}
\usepackage{longtable,tabu}
\usepackage[section]{placeins}
\usepackage{hyperref}
\hypersetup{%
  pdftitle   = {Supersymmetry of hyperbolic monopoles},
  pdfkeywords = {hyperbolic monopoles, supersymmetry, supersymmetric Yang--Mills--Higgs, curved space supersymmetry, pluricomplex geometry, hypercomplex geometry},
  pdfauthor  = {José Figueroa-O'Farrill, Moustafa Gharamti},
  pdfcreator = {\LaTeX\ with package \flqq hyperref\frqq}
}
\newcommand{\half}{\tfrac{1}{2}}

\newcommand{\fg}{\mathfrak{g}}

\newcommand{\Spin}{\mathrm{Spin}}
\ifx\Sp\UnDeFiNeD
\newcommand{\Sp}{\mathrm{Sp}}
\else
\renewcommand{\Sp}{\mathrm{Sp}}
\fi

\newcommand{\RR}{\mathbb{R}}
\newcommand{\CC}{\mathbb{C}}
\newcommand{\HH}{\mathbb{H}}

\newcommand{\ZZ}{\mathbb{Z}}
\newcommand{\II}{\mathbb{I}}

\newcommand{\eE}{\mathcal{E}}

\newcommand{\eG}{\mathcal{G}}

\newcommand{\eI}{\mathcal{I}}
\newcommand{\eJ}{\mathcal{J}}
\newcommand{\eK}{\mathcal{K}}
\newcommand{\eL}{\mathcal{L}}
\newcommand{\eM}{\mathcal{M}}

\newcommand{\eP}{\mathcal{P}}

\DeclareMathOperator{\Mat}{Mat}

\DeclareMathOperator{\Tr}{Tr}

\definecolor{orange}{rgb}{0.9,0.45,0}
\definecolor{green}{rgb}{0,0.5,0}
\newcommand{\MUNCH}[1]{\relax}

\allowdisplaybreaks[1]
\begin{document}
\title{Supersymmetry of hyperbolic monopoles}
\author[Figueroa-O'Farrill]{José Figueroa-O'Farrill}
\author[Gharamti]{Moustafa Gharamti}
\address{Maxwell and Tait Institutes, School of Mathematics, University of Edinburgh}
\thanks{EMPG-13-19}
\begin{abstract}
  We investigate what supersymmetry says about the geometry of the
  moduli space of hyperbolic monopoles.  We construct a
  three-dimensional supersymmetric Yang--Mills--Higgs theory on
  hyperbolic space whose half-BPS configurations coincide with
  (complexified) hyperbolic monopoles.  We then study the action of
  the preserved supersymmetry on the collective coordinates and show
  that demanding closure of the supersymmetry algebra constraints the
  geometry of the moduli space of hyperbolic monopoles, turning it
  into a so-called pluricomplex manifold, thus recovering a recent
  result of Bielawski and Schwachhöfer.
\end{abstract}
\maketitle
\tableofcontents

\section{Introduction}

BPS monopoles---that is, the solutions of the Bogomol'nyi
equation---have been under the microscope by mathematicians and
physicists for a long time.  This equation and its solutions can be
studied on any oriented riemannian 3-manifold, but they are
particularly interesting in euclidean and hyperbolic spaces.  One
inspiring observation about BPS monopoles in these spaces is that they
can be viewed as instantons in four-dimensional euclidean space left
invariant under the action of a one-parameter subgroup of isometries:
translations (resp. rotations) in the case of euclidean
(resp. hyperbolic) BPS monopoles.  Another way of saying this is that
the Bogomol'nyi equation results from the four-dimensional
self-duality equation by demanding independence on one of the
coordinates.

To begin with, consider the Bogomol'nyi equation in euclidean space
\begin{equation}
  \label{eq:mosbf}
  \nabla_A\phi=-\star F_A,
\end{equation}
where $\phi$ satisfies some suitable boundary conditions that make the
$L^2$ norm of $F_A$ finite and $\star$ is the Hodge operator of
$\RR^3$. For a detailed treatment of euclidean monopoles, one can
check \cite{Atiyah1988a, Manton2004, Shnir2005}.  The ingredients of
the Bogomol'nyi equation can be cast into a geometrical framework,
where $A$ can be viewed as a connection on a principal $G$-bundle $P$
over $\RR^3$ and $F_A$ as its curvature.  The Higgs field $\phi$ is a
section of the adjoint bundle $adP$ over $\RR^3$; that is, the
associated vector bundle to $P$ corresponding to the adjoint
representation of $G$ on its Lie algebra, and $\nabla_A$ is the
covariant derivative operator induced on $adP$.  A pair $(A,\phi)$
satisfying equation~\eqref{eq:mosbf} is what we call a euclidean
monopole.  If we now interpret $\phi$ as being the $x_4$ component of
the connection, then equation~\eqref{eq:mosbf} becomes the
self-duality Yang-Mills equation on $\RR^4$
\begin{equation}
  \label{eq:mossd}
  F_A=\star F_A,
\end{equation}
where all the fields are independent of the $x_4$ coordinate, and the
$\star$-operation is now with respect to the flat euclidean metric on
$\RR^4$.

For the case of hyperbolic monopoles we simply replace the euclidean
base space $\RR^3$ with hyperbolic space $H^3$.  To construct
hyperbolic monopoles from instantons, instead of considering
translationally invariant solutions of equation~\eqref{eq:mossd} we
will, however, look for rotationally invariant solutions
\cite{Atiyah1988}.  To be specific consider the flat euclidean metric
in $\RR^4$
\begin{equation}
  ds^2=dx^{2}_1+dx^{2}_2+dx^{2}_3+dx^{2}_4~.
\end{equation}
If we choose the rotations to be in the $(x_1,x_2)$-plane and we let
$r$ and $\theta$ be the polar coordinates in that plane, we have
\begin{equation}
  \begin{aligned}[m]
    ds^2 &= dr^2+r^2d\theta^2+dx^{2}_3+dx^{2}_4\\
    &= r^2 \left(d\theta^2+ \frac{dr^2+dx^{2}_3+dx^{2}_4}{r^2}\right)~.
  \end{aligned}
\end{equation}
The rotations now act simply as shifts in the angular variable
$\theta$.  This coordinate system is valid in the complement
$\RR^4\setminus\RR^2$ of the $x_1=x_2=0$ plane.  Inside the
parenthesis we recognise the metric on $S^1 \times H^3$, which is
therefore shown to be conformal to $\RR^4\setminus\RR^2$.

Now a wonderful fact about the self-duality equation is its conformal
invariance: the Hodge $\star$ is conformally invariant acting on
middle-dimensional forms in an even-dimensional manifold.  This allows
us to drop the conformal factor $r^2$ from the metric without altering
the equation.  If we now impose the condition that the gauge potential
$A$ is $S^1$ invariant, i.e., rotationally symmetric in the
$(x_1,x_2)$-plane, and if we define $A_\theta$=$\phi$, the
self-duality equation becomes the Bogomol'nyi equation on $H^3$.  The
Bogomol'nyi equation on $H^3$ is also given by
equation~\eqref{eq:mosbf} but with the $\star$-operation of $H^3$.
The first constructions of a monopole solution on hyperbolic space
were first given in \cite{Nash1986, Chakrabarti1986, Braam1989}.

A BPS monopole in hyperbolic space is labelled by a mass $m \in
\RR^{+}$ and a charge $k \in \ZZ^{+}$ given by
\begin{equation}
  \begin{aligned}[m]
    m &= \lim_{r\to\infty}|\phi(r)|\\
    k &=\lim_{r\to\infty}\frac{1}{4\pi m}\int_{H^3}tr(F_A\wedge \nabla_A\phi)~,
  \end{aligned}
\end{equation}
and it is known \cite{Sibner2012} that hyperbolic monopoles exist for
all values of $m$ and $k$. In contrast to the euclidean monopoles, $m$
cannot be rescaled to unity in the hyperbolic case, as the value of
$m$ affects the monopole solutions \cite{Braam1990}.  Alternatively,
one can normalise the mass to unity, but only at the price of
rescaling the hyperbolic metric to one of curvature $-1/m^2$.  The
rotationally invariant instanton on $\RR^4\setminus\RR^2$
corresponding to a hyperbolic monopole of charge $k$ and mass $m$ will
extend to a rotationally invariant instanton on all of $\RR^4$ if (and
only if) $m \in \ZZ$.

In \cite{Manton1982} Manton interpreted low energy dynamics of
monopoles as geodesic motion on the moduli space; that is, the space
of solutions up to gauge equivalence, and this ushered in an era of
much activity in the study of the geometry of the moduli space.  For
the case of euclidean monopoles, Atiyah and Hitchin showed in
\cite{Atiyah1988a} that the moduli space has a natural hyperkähler
metric and they found the explicit form of the metric for the moduli
space of charge $2$.  Moreover, the metric of the moduli space of well
separated monopoles was found in \cite{Manton1985}, where the
monopoles were treated as point particles carrying scalar, electric
and magnetic charges.

The hyperbolic case is much less understood.  In \cite{Atiyah1988},
where Atiyah introduced hyperbolic monopoles, he writes:
\begin{quote}
  Moreover, by varying the curvature of hyperbolic space and letting
  it tend to zero, the euclidean case appears as a natural limit of
  the hyperbolic case. While the details of this limiting procedure
  are a little delicate, and need much more careful examination than I
  shall give here, it seems reasonable to conjecture that the moduli
  of monopoles remains unaltered by passing to the limit.
\end{quote}
Atiyah also showed \cite{MR763752} that the moduli space $\eM_{k,m}$
of hyperbolic monopoles of charge $k$ and mass $m$ can be identified
with the space of rational maps of the form
\begin {equation}
  \frac{a_1z^{k-1}+a_2z^{k-2}+\dots+a_k}{z^k+b_1z^{k-1}+\dots+b_k} \quad\text{with $k\geq 1$,}
\end{equation}
where the polynomials in the numerator and denominator are relatively
prime.  Since the $a_1,\dots,a_k,b_1,\dots,b_k$ are complex numbers,
the moduli space has real dimension $4k$.

Most of the progress in the study of hyperbolic monopoles was focused
on finding methods of constructing multimonopole solutions, either by
building a hyperbolic version of the Nahm transform \cite{Braam1990,
  Ward1998, Manton:2012xn} or by studying the spectral curves
associated with hyperbolic monopoles \cite{Murray1996, Murray2000,
  Norbury2007}.  Progress on the geometry of the moduli space was
hindered by the early realisation \cite{Braam1990} that the natural
$L^2$ metric, which in the euclidean case induces upon reduction a
hyperkähler metric on the moduli space, does not converge in the case
of hyperbolic monopoles, suggesting that the geometry of the moduli
space is not in fact riemannian.  Nevertheless, Hitchin
\cite{Hitchin1996} constructed a family $g_m$ of self-dual Einstein
metrics on the moduli space of centered hyperbolic monopoles with mass
$m \in \ZZ$, which in the flat limit $m\to \infty$ recovers the
Atiyah--Hitchin metric.  It is an interesting open question to relate
Hitchin's construction to the physics of hyperbolic monopoles.

The situation has changed dramatically in recent times due to the
seminal work of Bielawski and Schwachhöfer, based on earlier work of
O.~Nash \cite{Nash2007}.  Nash used a new twistorial construction of
$\eM_{k,m}$ to show that the complexification of the real geometry of
the moduli space of hyperbolic monopoles is similar in some respects
to the complexification of a hyperkähler geometry.  Building on that
work, Bielawski and Schwachhöfer \cite{Bielawski2013} identified the
real geometry of the moduli space of hyperbolic monopoles as
``pluricomplex geometry'', which is equivalent to saying that there is
a $\CC$-linear hypercomplex structure on the complexification
$T_\CC\eM_{k,m}$ of the tangent bundle to the moduli space.  Later in
\cite{Bielawski:2012np} Bielawski and Schwachhöfer studied the
euclidean limit of the pluricomplex moduli space of hyperbolic
monopoles, and showed that in the limit one recovers an enhanced
hyperkähler geometry, richer by an additional complex structure.

The fact that BPS monopoles saturate the Bogomol'nyi bound suggests
that monopoles are supersymmetric in nature and in this paper we will
exhibit this in detail for the case of the hyperbolic monopoles.
Similar results for the case of euclidean monopoles were obtained in
\cite{A.DAdda1978, Harvey1993, Gauntlett1994} among others.  The aim
of this paper is thus to show that the pluricomplex nature of the
moduli space of hyperbolic monopoles is a natural consequence of
supersymmetry.  One novel aspect of our construction is that the
contraints coming from supersymmetry are imposed by demanding the
closure of the supersymmetry algebra and not the invariance of the
effective action for the moduli, which does not exist due to the lack
of convergence of the $L^2$ metric.  This is reminiscent of the
results of Stelle and Van Proeyen \cite{Stelle:2003rr} on Wess--Zumino
models without an action functional, in which the geometry is relaxed
from Kähler to complex flat.  In fact, morally one could say that
pluricomplex is to hyperkähler what complex flat is to Kähler.

The paper is organised as follows.  In
Section~\ref{sec:hyperb-supersymm-yan} we construct a supersymmetric
Yang--Mills--Higgs theory in hyperbolic space by starting with
supersymmetric Yang--Mills theory on Minkowski spacetime,
euclideanising to a supersymmetric Yang--Mills theory on $\RR^4$,
reducing to $\RR^3$ and deforming to a supersymmetric theory on $H^3$.
In Section~\ref{sec:moduli-space-bps} we show that the hyperbolic
monopoles coincide with the configurations which preserve precisely
one half of the supersymmetry.  We also start the analysis of the
moduli space by studying the linearisation of the Bogomol'nyi equation
and identifying the bosonic and fermionic zero modes and how the
unbroken supersymmetry relates them.  A possibly surprising result is
the fact that supersymmetry suggests a small modification of the Gauss
law constraint, which depends explicitly on the hyperbolic curvature.
Finally in Section~\ref{sec:geom-moduli-space} we linearise the
unbroken supersymmetry and demanding the on-shell closure of the
supersymmetry algebra will yield the conditions satisfied by the
geometry of the moduli space.  The paper ends with an appendix on the
Frölicher--Nijenhuis bracket of two endomorphisms.

\label{sec:introduction}

\section{Hyperbolic supersymmetric Yang--Mills--Higgs}
\label{sec:hyperb-supersymm-yan}

The purpose of this section is to describe a construction of
supersymmetric theories in hyperbolic space by the following
procedure: start with supersymmetric Yang--Mills in Minkowski
spacetime, euclideanise à la van Nieuwenhuizen--Waldron
\cite{Nieuwenhuizen1996}, reduce to $\RR^3$ and deform to a theory on
$H^3$.  The euclideanisation will require complexifying the fields in
the theory.

\subsection{Off-shell supersymmetry in euclidean 4-space}

The first step has been done in \cite{Nieuwenhuizen1996}, except that
we expect that auxiliary fields should play an important rôle and thus
must promote the theory to one with off-shell closure of supersymmetry
(up to possibly gauge transformations).

The euclidean supersymmetric Yang--Mills action in $\RR^4$ is obtained
by integrating the lagrangian density
\begin{equation}
  \label{eq:esymr4}
  \eL^{(4)} = - \Tr \chi_R^\dagger \slashed{D} \psi_L - \tfrac14 \Tr F^2~,
\end{equation}
where $\Tr$ denotes an ad-invariant inner product on the Lie algebra
$\fg$ of the gauge group $G$, and where the subscripts $L,R$ denote
the projections
\begin{equation}
  \psi_L = \half (\II + \gamma^5) \psi \qquad\text{and}\qquad \chi_R^\dagger = \half \chi^\dagger (\II-\gamma^5)~,
\end{equation}
where $\gamma^5 = \gamma^1 \gamma^2 \gamma^3 \gamma^4$, where
$\gamma^\mu \gamma^\nu = \gamma^{\mu\nu} + \delta^{\mu\nu}$.  This
means that that $(\gamma^5)^2 = 1$.  We can raise and lower indices
with impunity, since the metric is $\delta_{\mu\nu}$.  The action
defined by $\eL^{(4)}$ is invariant under gauge transformations, which
infinitesimally take the form
\begin{equation}
  \delta_\Lambda \psi_L = [\Lambda,\psi_L] \qquad   \delta_\Lambda \chi_R^\dagger = [\Lambda,\chi_R^\dagger] \qquad\text{and}\qquad \delta_\Lambda A_\mu = - D_\mu \Lambda = - \partial_\mu \Lambda + [\Lambda, A_\mu]~,
\end{equation}
with $\Lambda \in C^\infty(\RR^4;\fg)$.  Furthermore it is invariant
under the supersymmetry transformations
\begin{equation}
  \begin{aligned}[m]
    \delta_\varepsilon \psi_L &= \half \gamma^{\mu\nu}F_{\mu\nu} \varepsilon_L\\
    \delta_\varepsilon \chi_R^\dagger &= - \half \varepsilon_R^\dagger \gamma^{\mu\nu}F_{\mu\nu}\\
    \delta_\varepsilon A_\mu &= - \varepsilon_R^\dagger \gamma_\mu \psi_L + \chi_R^\dagger \gamma_\mu \varepsilon_L~,
 \end{aligned}
\end{equation}
where $\varepsilon_L$ and $\varepsilon_R^\dagger$ are constant spinor
parameters of the indicated chirality.  Since $\varepsilon_L$ and
$\varepsilon_R^\dagger$ are independent, we actually have two
supersymmetry variations, which we will denote $\delta_L$ and
$\delta_R$ and leave the parameter unspecified when there is no danger
of confusion.  In this notation we have
\begin{equation}
  \begin{aligned}[m]
    \delta_L \psi_L &= \half \gamma^{\mu\nu}F_{\mu\nu} \varepsilon_L\\
    \delta_L \chi_R^\dagger &= 0\\
    \delta_L A_\mu &= \chi_R^\dagger \gamma_\mu \varepsilon_L
 \end{aligned}
\qquad\qquad
  \begin{aligned}[m]
    \delta_R \psi_L &= 0\\
    \delta_R \chi_R^\dagger &= - \half \varepsilon_R^\dagger \gamma^{\mu\nu}F_{\mu\nu}\\
    \delta_R A_\mu &= - \varepsilon_R^\dagger \gamma_\mu \psi_L~.
 \end{aligned}
\end{equation}

Notice that if $\delta'_L$ is defined as $\delta_L$ but with a
different supersymmetry parameter, say $\varepsilon'_L$, then on the
gauge field $[\delta_L,\delta'_L] A_\mu = 0$, and similarly
$[\delta_R,\delta'_R] A_\mu = 0$.  On the fermion, however, this will
not be true off-shell and it is for that reason that we will introduce
an auxiliary field.  Indeed, one finds
\begin{equation}
  [\delta_L,\delta'_L] \psi_L = \delta_L (\half \gamma^{\mu\nu}F_{\mu\nu} \varepsilon'_L) - \delta'_L (\half \gamma^{\mu\nu}F_{\mu\nu} \varepsilon_L)~.
\end{equation}
Using that
\begin{equation}
  \delta_L F_{\mu\nu} = D_\mu \delta_L A_\nu - D_\nu \delta_L A_\mu = D_\mu(\chi_R^\dagger \gamma_\nu \varepsilon_L) - D_\nu(\chi_R^\dagger \gamma_\mu \varepsilon_L)~,
\end{equation}
whence
\begin{equation}
  \label{eq:dldl1}
  [\delta_L,\delta'_L] \psi_L = D_\mu \chi_R^\dagger \gamma_\nu \varepsilon_L \gamma^{\mu\nu} \varepsilon'_L - D_\mu \chi_R^\dagger \gamma_\nu \varepsilon'_L \gamma^{\mu\nu} \varepsilon_L~,
\end{equation}
where we have used that $\gamma_\mu^\dagger = \gamma_\mu$ and also
that $(\chi^\dagger\psi)^\dagger = + \psi^\dagger \chi$ for
anticommuting spinors. (One might think that the $+$ sign violates the
sign rule, but it does not because $\psi$ and $\psi^\dagger$ are
independent fields, etc.)

In order to further manipulate the right-hand side of
$[\delta_L,\delta'_L] \psi_L$ we must make use of a Fierz identity.
The basic Fierz identity in $\RR^4$ for anticommuting spinors is given
by
\begin{equation}
  \label{eq:fierz}
  \psi\chi^\dagger = -\tfrac14 \chi^\dagger \psi \II - \tfrac14 \chi^\dagger\gamma^5\psi \gamma_5 - \tfrac14 \chi^\dagger \gamma^\mu \psi \gamma_\mu + \tfrac14 \chi^\dagger \gamma^\mu \gamma^5 \psi \gamma_\mu \gamma_5 + \tfrac18  \chi^\dagger \gamma^{\mu\nu} \psi \gamma_{\mu\nu}~.
\end{equation}
Two special cases will play a rôle in what follows:
\begin{equation}
  \label{eq:fierzLR}
  \psi_L \chi_R^\dagger = -\half \chi_R^\dagger \gamma^\mu \psi_L \gamma_\mu P_R~,
\end{equation}
and
\begin{equation}
  \label{eq:fierzRR}
  \psi_R \chi_R^\dagger = -\tfrac12 \chi_R^\dagger \psi_R P_R - \tfrac18 \chi_R^\dagger \gamma^{\mu\nu} \psi_R \gamma_{\mu\nu}~,
\end{equation}
where $P_R = \half (\II - \gamma_5)$.  Of course, for commuting
spinors, we simply flip all signs in the right-hand side.

Using the Fierz formula \eqref{eq:fierzLR}, we may rewrite
\begin{equation}
  [\delta_L,\delta'_L] \psi_L = -\half D_\mu \chi_R^\dagger \gamma^\sigma \varepsilon'_L \gamma^{\mu\nu} \gamma_\sigma \gamma_\nu \varepsilon_L + \half D_\mu \chi_R^\dagger \gamma^\sigma \varepsilon_L \gamma^{\mu\nu} \gamma_\sigma \gamma_\nu \varepsilon'_L~.
\end{equation}
Using that $\gamma^{\mu\nu} \gamma_\sigma \gamma_\nu =
-\gamma_{\mu\sigma} - 3 \delta_{\mu\sigma}$, we rewrite
\begin{equation}
  [\delta_L,\delta'_L] \psi_L = \tfrac32 \chi_R^\dagger
\overleftarrow{\slashed{D}}\varepsilon'_L \varepsilon_L - \tfrac32
\chi_R^\dagger \overleftarrow{\slashed{D}}\varepsilon_L \varepsilon'_L
+ \half D_\mu\chi_R^\dagger \gamma_\nu \varepsilon'_L
\gamma^{\mu\nu}\varepsilon_L - \half D_\mu\chi_R^\dagger \gamma_\nu
\varepsilon_L \gamma^{\mu\nu}\varepsilon'_L~.
\end{equation}
Comparing with equation~\eqref{eq:dldl1} we see that
\begin{equation}
  \mu \chi_R^\dagger \gamma_\nu \varepsilon_L \gamma^{\mu\nu}
\varepsilon'_L - D_\mu \chi_R^\dagger \gamma_\nu \varepsilon'_L
\gamma^{\mu\nu} \varepsilon_L = \chi_R^\dagger
\overleftarrow{\slashed{D}}\varepsilon'_L \varepsilon_L -
\chi_R^\dagger \overleftarrow{\slashed{D}}\varepsilon_L
\varepsilon'_L~,
\end{equation}
whence, in summary,
\begin{equation}
  [\delta_L,\delta'_L] \psi_L = \chi_R^\dagger
\overleftarrow{\slashed{D}}\varepsilon'_L \varepsilon_L -
\chi_R^\dagger \overleftarrow{\slashed{D}}\varepsilon_L
\varepsilon'_L~,
\end{equation}
which vanishes for all $\varepsilon_L,\varepsilon'_L$ if and only if
$\chi_R^\dagger \overleftarrow{\slashed{D}} = 0$, which is the field
equation for $\chi_R^\dagger$.  This suggests introducing an auxiliary
field, historically denoted by $D$ (and who are we to challenge
tradition?!), and modifying the supersymmetry variation of $\psi_L$ by
a term proportional to $D$, namely
\begin{equation}
  \delta_L \psi_L = D \varepsilon_L + \half \gamma^{\mu\nu} F_{\mu\nu} \varepsilon_L~.
\end{equation}
Now, we see that
\begin{equation}
  [\delta_L,\delta'_L] \psi_L = (\delta_L D -\chi_R^\dagger \overleftarrow{\slashed{D}}\varepsilon_L) \varepsilon'_L - (\delta'_L D - \chi_R^\dagger \overleftarrow{\slashed{D}}\varepsilon'_L) \varepsilon_L~,
\end{equation}
whence we deduce that if we set
\begin{equation}
  \delta_L D  = \chi_R^\dagger \overleftarrow{\slashed{D}}\varepsilon_L = D_\mu \chi_R^\dagger \gamma^\mu \varepsilon_L
\end{equation}
then $[\delta_L,\delta'_L] \psi_L = 0$.  But now we have to check that $[\delta_L,\delta'_L] D = 0$ as well:
\begin{equation}
  \label{eq:dldlD}
  \begin{split}
    [\delta_L,\delta'_L] D &= \delta_L (D_\mu \chi_R^\dagger \gamma^\mu \varepsilon'_L) - \delta'_L (D_\mu \chi_R^\dagger \gamma^\mu \varepsilon_L) \\
    &= [\delta_L A_\mu, \chi_R^\dagger] \gamma^\mu \varepsilon'_L - [\delta'_L A_\mu, \chi_R^\dagger] \gamma^\mu \varepsilon_L\\
    &= 2 [\chi_R^\dagger \gamma_\mu \varepsilon_L, \chi_R^\dagger \gamma^\mu \varepsilon'_L]~,
  \end{split}
\end{equation}
where we have used that $\delta_L \chi_R^\dagger = 0$.  We now use the Fierz identity~\eqref{eq:fierzLR}
and (in matrix notation) rewrite
\begin{equation}
  \begin{split}
    [\delta_L,\delta'_L] D &= 2 \chi_R^\dagger \gamma_\mu \varepsilon_L \chi_R^\dagger \gamma^\mu \varepsilon'_L - 2 \chi_R^\dagger \gamma_\mu \varepsilon'_L \chi_R^\dagger \gamma^\mu \varepsilon_L\\
    &= - \chi_R^\dagger \gamma_\mu \gamma_\nu \gamma^\mu \varepsilon'_L \chi_R^\dagger \gamma^\nu \varepsilon_L + 
    \chi_R^\dagger \gamma_\mu \gamma_\nu \gamma^\mu \varepsilon_L \chi_R^\dagger \gamma^\nu \varepsilon'_L\\
    &= 2 \chi_R^\dagger \gamma_\nu \varepsilon'_L \chi_R^\dagger \gamma^\nu \varepsilon_L -
    2 \chi_R^\dagger \gamma_\nu \varepsilon_L \chi_R^\dagger \gamma^\nu \varepsilon'_L\\
    &= 2 [\chi_R^\dagger \gamma_\mu \varepsilon'_L, \chi_R^\dagger \gamma^\mu \varepsilon_L]~,
  \end{split}
\end{equation}
which is to be compared with equation~\eqref{eq:dldlD}, from where we see that indeed $[\delta_L,\delta'_L] D = 0$.

In a similar way we work out $\delta_R D$ by the requirement that $[\delta_R,\delta'_R]\chi_R^\dagger = 0$.  Let $\alpha$ be a number to be determined and let
\begin{equation}
  \delta_R \chi_R^\dagger = \alpha D \varepsilon_R^\dagger - \half \varepsilon_R^\dagger \gamma^{\mu\nu} F_{\mu\nu}~.
\end{equation}
Then
\begin{equation}
\label{eq:drdrchi}
  \begin{split}
    [\delta_R,\delta'_R]\chi_R^\dagger &= \delta_R \left(\alpha D \varepsilon_R^\dagger - \half \varepsilon'_R{}^\dagger \gamma^{\mu\nu} F_{\mu\nu}\right) - \delta'_R \left(\alpha D \varepsilon_R^\dagger - \half \varepsilon_R^\dagger \gamma^{\mu\nu} F_{\mu\nu}\right)\\
    &=\alpha \delta_R D \varepsilon'_R{}^\dagger + \varepsilon_R^\dagger \gamma_\nu D_\mu \psi_L \varepsilon'_R{}^\dagger \gamma^{\mu\nu} -  (\varepsilon_R \leftrightarrow \varepsilon'_R) ~.
  \end{split}
\end{equation}
We use the Fierz identity~\eqref{eq:fierzLR}
\begin{equation}
  D_\mu \psi_L \varepsilon'_R{}^\dagger = - \half \varepsilon'_R{}^\dagger \gamma^\sigma D_\mu\psi_L \gamma_\sigma P_R
\end{equation}
to rewrite
\begin{equation}
  \begin{split}
    [\delta_R,\delta'_R]\chi_R^\dagger &= \alpha \delta_R D \varepsilon'_R{}^\dagger - \half \varepsilon'_R{}^\dagger \gamma^\sigma D_\mu \psi_L \varepsilon_R^\dagger \gamma_\nu \gamma_\sigma \gamma^{\mu\nu} - (\varepsilon_R \leftrightarrow \varepsilon'_R)~.
  \end{split}
\end{equation}
We now use that $\gamma_\nu \gamma_\sigma \gamma^{\mu\nu} = - \gamma_{\mu\sigma} + 3 \delta_{\mu\sigma}$ to rewrite the above equation as
\begin{equation}
    \begin{split}
    [\delta_R,\delta'_R]\chi_R^\dagger &= \alpha \delta_R D \varepsilon'_R{}^\dagger + \half \varepsilon'_R{}^\dagger \gamma_\sigma D_\mu \psi_L \varepsilon_R^\dagger \gamma^{\mu\sigma} - \tfrac32  \varepsilon'_R{}^\dagger \slashed{D} \psi_L \varepsilon_R^\dagger - (\varepsilon_R \leftrightarrow \varepsilon'_R)~.
  \end{split}
\end{equation}
Comparing with equation~\eqref{eq:drdrchi}, we see that
\begin{equation}
  \varepsilon'_R{}^\dagger \gamma_\sigma D_\mu \psi_L \varepsilon_R^\dagger \gamma^{\mu\sigma} - (\varepsilon_R \leftrightarrow \varepsilon'_R) = \varepsilon'_R{}^\dagger \slashed{D} \psi_L \varepsilon_R^\dagger - (\varepsilon_R \leftrightarrow \varepsilon'_R)~,
\end{equation}
whence finally
\begin{equation}
  [\delta_R,\delta'_R]\chi_R^\dagger = \left(\alpha \delta_R D + \varepsilon_R^\dagger \slashed{D} \psi_L \right) \varepsilon'_R{}^\dagger  - (\varepsilon_R \leftrightarrow \varepsilon'_R)~,
\end{equation}
which vanishes provided that
\begin{equation}
  \delta_R D = -\frac1\alpha \varepsilon_R^\dagger \slashed{D} \psi_L ~.
\end{equation}
As before, one checks that $[\delta_R,\delta'_R]D = 0$.

We fix $\alpha$ by closing the supersymmetry algebra on the gauge field: we expect that it should close to a translation up to a gauge transformation.  Indeed,
\begin{equation}
  \begin{split}
    [\delta_L,\delta_R] A_\mu &= \delta_L (-\varepsilon_R^\dagger \gamma_\mu \psi_L)- \delta_R (\chi_R^\dagger \gamma_\mu     \varepsilon_L)\\
    &= - \varepsilon_R^\dagger \gamma_\mu \left(D + \half \gamma^{\nu\rho} F_{\nu\rho}\right) \varepsilon_L 
    - \varepsilon_R^\dagger \left(\alpha D - \half \gamma^{\nu\rho} F_{\nu\rho}\right) \gamma_\mu \varepsilon_L\\
    &= - (1+\alpha)\varepsilon_R^\dagger \gamma_\mu \varepsilon_L D - \half \varepsilon_R^\dagger \left( \gamma_\mu \gamma^{\nu\rho} - \gamma^{\nu\rho} \gamma_\mu \right) \varepsilon_L  F_{\nu\rho}~,
  \end{split}
\end{equation}
whence we see that $\alpha = -1$ and using that $[\gamma_\mu,\gamma^{\nu\rho}] = 2 \delta_\mu^\nu \gamma^\rho - 2 \delta_\mu^\rho \gamma^\nu$, we rewrite
\begin{equation}
  \begin{split}
    [\delta_L,\delta_R] A_\mu &= 2 \varepsilon_R^\dagger \gamma^\rho \varepsilon_L  F_{\rho\mu}\\
    &= 2 \varepsilon_R^\dagger \gamma^\rho \varepsilon_L  (\partial_\rho A_\mu - \partial_\mu A_\rho + [A_\rho,A_\mu])\\
    &= \xi^\rho \partial_\rho A_\mu - D_\mu \Lambda~,
  \end{split}
\end{equation}
where $\xi^\rho = 2 \varepsilon_R^\dagger \gamma^\rho \varepsilon_L$ and $\Lambda = \xi^\rho A_\rho$.

In a similar way, one shows that the algebra closes as expected also on $\psi_L$, $\chi_R^\dagger$ and $D$.  Indeed, on $\psi_L$ one has
\begin{equation}
  \begin{split}
    [\delta_L,\delta_R] \psi_L &= -\delta_R (D \varepsilon_L + \half \gamma^{\mu\nu} F_{\mu\nu}\varepsilon_L)\\
    &= - \varepsilon_R^\dagger \slashed{D}\psi_L \varepsilon_L - \gamma^{\nu\mu} \varepsilon_R^\dagger \gamma_\nu D_\mu \psi_L \varepsilon_L\\
    &= -\gamma^\nu \gamma^\mu \varepsilon_L \varepsilon_R^\dagger\gamma_\nu D_\mu \psi_L~,
  \end{split}
\end{equation}
which upon using the Fierz identity \eqref{eq:fierz} for $\varepsilon_L \varepsilon_R^\dagger$ becomes
\begin{equation}
   [\delta_L,\delta_R] \psi_L = \half \varepsilon_R^\dagger \gamma^\rho \varepsilon_L \gamma^\nu
    \gamma^\mu \gamma_\rho \gamma_\nu D_\mu \psi_L~.
\end{equation}
Now, we use that $\gamma^\nu \gamma_{\mu\rho} \gamma_\nu = 0$ in four dimensions in order to rewrite this as
\begin{equation}
   [\delta_L,\delta_R] \psi_L = 2  \varepsilon_R^\dagger \gamma^\mu \varepsilon_L D_\mu \psi_L = \xi^\mu \partial_\mu \psi_L + [\Lambda, \psi_L]~,
\end{equation}
as expected.  The calculation for $[\delta_L,\delta_R] \chi_R^\dagger$ is similar.  Finally, we check closure on $D$:
\begin{equation}
  \begin{split}
    [\delta_L,\delta_R] D &= \delta_L (\varepsilon_R^\dagger \slashed{D} \psi_L) - \delta_R (\chi_R^\dagger \overleftarrow{\slashed{D}}\varepsilon_L)\\
    &= \varepsilon_R^\dagger \gamma_\mu [\chi_R^\dagger \gamma^\mu \varepsilon_L, \psi_L] + \varepsilon_R^\dagger \slashed{D}(D \varepsilon_L + \half \gamma^{\mu\nu} F_{\mu\nu} \varepsilon_L)\\
    &\qquad + (\varepsilon_R^\dagger D + \half \varepsilon_R^\dagger \gamma^{\mu\nu} F_{\mu\nu}) \overleftarrow{\slashed{D}} \varepsilon_L + [\varepsilon_R^\dagger \gamma_\mu \psi_L, \chi_R^\dagger] \gamma^\mu \varepsilon_L\\
    &= \varepsilon_R^\dagger \gamma^\rho (D_\rho D + \half \gamma^{\mu\nu} D_\rho F_{\mu\nu}) \varepsilon_L + 
    \varepsilon_R^\dagger (D_\rho D + \half \varepsilon_R^\dagger \gamma^{\mu\nu} D_\rho F_{\mu\nu}) \gamma^\rho \varepsilon_L\\
    &= 2 \varepsilon_R^\dagger \gamma^\rho D_\rho D \varepsilon_L + \half \varepsilon_R^\dagger (\gamma^\rho \gamma^{\mu\nu} + \gamma^{\mu\nu} \gamma^\rho) D_\rho F_{\mu\nu} \varepsilon_L~.
  \end{split}
\end{equation}
Using that $\gamma^\rho \gamma^{\mu\nu} + \gamma^{\mu\nu} \gamma^\rho = 2 \gamma^{\rho\mu\nu}$ and the Bianchi identity $D_{[\rho} F_{\mu\nu]}= 0$, we conclude that
\begin{equation}
  [\delta_L,\delta_R] D = 2 \varepsilon_R^\dagger \gamma^\rho D_\rho D \varepsilon_L = \xi^\rho \partial_\rho D + [\Lambda, D]~,
\end{equation}
as desired.

In summary, the following supersymmetry transformations
\begin{equation}
  \label{eq:esusy4d}
  \begin{aligned}[m]
    \delta_L A_\mu &= \chi_R^\dagger \gamma_\mu \varepsilon_L \\
    \delta_L \psi_L &= D \varepsilon_L + \half \gamma^{\mu\nu} F_{\mu\nu} \varepsilon_L\\
    \delta_L \chi_R^\dagger &= 0\\
    \delta_L D &= \chi_R^\dagger \overleftarrow{\slashed{D}}\varepsilon_L
 \end{aligned}
  \qquad\qquad
  \begin{aligned}[m]
    \delta_R A_\mu &= - \varepsilon_R^\dagger \gamma_\mu \psi_L \\
    \delta_R \psi_L &= 0\\
    \delta_R \chi_R^\dagger &= - \varepsilon_R^\dagger D - \half \varepsilon_R^\dagger \gamma^{\mu\nu} F_{\mu\nu}\\
    \delta_R D &= \varepsilon_R^\dagger \slashed{D} \psi_L
 \end{aligned}
\end{equation}
obey
\begin{equation}
  [\delta_L,\delta'_L] = 0\qquad [\delta_R,\delta'_R] = 0 \qquad\text{whereas}\qquad [\delta_L,\delta_R] = \eL_\xi + \delta_\Lambda^{\text{gauge}}~,
\end{equation}
where $\xi^\mu = 2 \varepsilon_R^\dagger \gamma^\mu \varepsilon_L$ and $\Lambda = \xi^\mu A_\mu$.

The action given by the lagrangian~\eqref{eq:esymr4} is not invariant under the supersymmetry transformations in \eqref{eq:esusy4d} unless we also add a term depending on the auxiliary field.  Indeed, the invariant action is given by
\begin{equation}
  \label{eq:esymr4aux}
  \eL^{(4)} = - \Tr \chi_R^\dagger \slashed{D} \psi_L - \tfrac14 \Tr F^2 - \half \Tr D^2~.
\end{equation}
It should be remarked that the euclideanisation has in fact complexified the fields in the original Yang--Mills theory.  Indeed, the spinor representation in euclidean signature is not of real type, as it is in lorentzian signature and the supersymmetry transformations further force the bosonic fields to be complex as well.

We may promote this action to an arbitrary riemannian 4-manifold simply by covariantising the derivatives, so that $D_\mu$ now also contains the spin connection.  Doing so and taking $\varepsilon_L$ and $\varepsilon_R^\dagger$ to be spinor fields, we find that
\begin{equation}
  \label{eq:dlesymr4aux}
  \delta_L \eL^{(4)} = -\nabla_\mu \Tr \chi_R^\dagger \gamma_\nu \varepsilon_L (D g^{\mu\nu} + F^{\mu\nu}) - \half \Tr \chi_R^\dagger \gamma^\rho \gamma^{\mu\nu} F_{\mu\nu} \nabla_\rho \varepsilon_L~,
\end{equation}
and
\begin{equation}
  \label{eq:dresymr4aux}
  \delta_R \eL^{(4)} = \half \nabla_\rho \Tr F_{\mu\nu} \varepsilon_R^\dagger \gamma^{\mu\nu\rho} \psi_L - \half \Tr \nabla_\rho \varepsilon_R^\dagger \gamma^{\mu\nu} \gamma^\rho F_{\mu\nu} \psi_L~,
\end{equation}
from where we see that if $\varepsilon_L$ and $\varepsilon_R^\dagger$ are not parallel, the action is not invariant.  This will be remedied for the dimensionally reduced action in three dimensions by adding further terms in the action provided that $\varepsilon_L$ and $\varepsilon_R^\dagger$ are Killing spinors.

\subsection{Reduction to euclidean 3-space}

The spin group in four dimensions is $\Spin(4) \cong \Spin(3)\times
\Spin(3)$.  The spin group in three dimensions is $\Spin(3)$ and
embeds in $\Spin(4)$ as the diagonal $\Spin(3)$ in $\Spin(3)\times
\Spin(3)$.  Therefore in three dimensions there is no distinction
between $L$ and $R$ spinors.  We reduce to three dimensions along the
fourth coordinate, whence we assume that $\partial_4 = 0$ on all
fields and parameters.

We take the following explicit realisation for the four-dimensional gamma matrices:
\begin{equation}
  \gamma_j =
  \begin{pmatrix}
    0 & -i \sigma^j \\ i \sigma^j & 0
  \end{pmatrix}
  \qquad
  \gamma_4 =
  \begin{pmatrix}
    0 & \II \\ \II & 0
  \end{pmatrix}
  \qquad\text{and hence}\qquad
  \gamma_5 = \gamma_1\gamma_2\gamma_3\gamma_4 =
  \begin{pmatrix}
    \II & 0 \\ 0 & -\II
  \end{pmatrix}~.
\end{equation}
This means that we can take $\psi_L = \begin{pmatrix}\psi \\ 0\end{pmatrix}$ and $\chi_R^\dagger =
\begin{pmatrix} 0 & \chi^\dagger \end{pmatrix}$.  The basic Fierz identity for anticommuting spinors in three dimensions is
\begin{equation}
  \label{eq:fierz3d}
  \psi \chi^\dagger = - \half \chi^\dagger \psi - \half \chi^\dagger \sigma^j \psi \sigma^j~.
\end{equation}
The gauge field decomposes as $A_\mu \rightsquigarrow (A_i, \phi)$.  The supersymmetry parameters $\varepsilon_L$ and $\varepsilon_R^\dagger$ also decompose as $\psi_L$ and $\chi_R^\dagger$ do: $\varepsilon_L = \begin{pmatrix}\epsilon_L \\ 0\end{pmatrix}$ and $\varepsilon_R^\dagger = \begin{pmatrix} 0 & \epsilon_R^\dagger \end{pmatrix}$.  In terms of the three-dimensional quantities we have the following supersymmetry transformations:
\begin{equation}
  \begin{aligned}
    \delta_L A_i &= i \chi^\dagger \sigma_i \epsilon_L\\
    \delta_L \phi &= \chi^\dagger \epsilon_L\\
    \delta_L \chi^\dagger &= 0\\
    \delta_L D &= i \chi^\dagger \overleftarrow{\slashed{D}} \epsilon_L + [\phi,\chi^\dagger\epsilon_L]\\
    \delta_L \psi &= D \epsilon_L + \tfrac{i}2 \varepsilon_{ijk} F^{ij}\sigma^k \epsilon_L - i D_i\phi\sigma^i \epsilon_L
  \end{aligned}
 \qquad
  \begin{aligned}
    \delta_R A_i &= - i \epsilon_R^\dagger \sigma_i \psi\\
    \delta_R \phi &= - \epsilon_R^\dagger \psi\\
    \delta_R \chi^\dagger &= -D \epsilon_R^\dagger - \tfrac{i}2 \varepsilon_{ijk} F^{ij}\epsilon_R^\dagger \sigma^k - i \epsilon_R^\dagger \sigma^i D_i\phi\\
    \delta_R D &=  i \epsilon_R^\dagger \slashed{D} \psi + \epsilon_R^\dagger [\phi,\psi]\\
    \delta_R \psi &= 0~,
  \end{aligned}
\end{equation}
where now
\begin{equation}
  [\delta_L,\delta'_L] = 0 = [\delta_R,\delta'_R] \qquad\text{and}\qquad [\delta_L,\delta_R] = \eL_\xi + \delta^{\text{gauge}}_\Lambda~,
\end{equation}
with $\xi^i = 2 i \epsilon_R^\dagger \sigma^i \epsilon_L$ and $\Lambda = \xi^i A_i + 2 \epsilon_R^\dagger \epsilon_L \phi$.

The reduction of the action~\eqref{eq:esymr4aux} to three dimensions is
\begin{equation}
  \label{eq:esymr3aux}
  \eL^{(3)} = -i \Tr \chi^\dagger \slashed{D} \psi - \Tr \chi^\dagger [\phi,\psi] - \tfrac14 \Tr F^2 - \half \Tr |D\phi|^2 - \tfrac12 \Tr D^2~,
\end{equation}
where $\slashed{D} = \sigma^i D_i$, $F^2 = F_{ij} F^{ij}$ and $|D\phi|^2 = D_i\phi D^i\phi$.  It can again be suitably covariantised to define it on a riemannian 3-manifold.  Its variation under supersymmetry can be read off from equations~\eqref{eq:dlesymr4aux} and \eqref{eq:dresymr4aux}.  Doing so, one finds
\begin{equation}
  \label{eq:dlesymr3aux}
  \delta_L \eL^{(3)} = -i \nabla_i \Tr \chi^\dagger \left( \sigma^i D + \sigma_j F^{ij} -i D^i\phi \right)\epsilon_L + \Tr \chi^\dagger \sigma^i  \sigma^\ell \left(\half \varepsilon_{jk\ell} F^{jk} - D_\ell\phi \right) \nabla_i\epsilon_L
\end{equation}
and
\begin{equation}
  \label{eq:dresymr3aux}
  \delta_R \eL^{(3)} = \nabla_i \Tr \varepsilon^{ijk} \epsilon_R^\dagger \left(-\half F_{jk} +i D_j\phi \sigma_k\right) \psi + Tr \nabla_i \epsilon_R^\dagger \left(\half \varepsilon_{jk\ell} F^{jk} + D_\ell \phi\right) \sigma^\ell \sigma^i \psi~.
\end{equation}

\subsection{Deforming to curved space}

We now wish to improve the action $\eL^{(3)}$ and the supersymmetry
transformations of the fermions and the auxiliary field in order for
the new $\eL^{(3)}$ to transform into a total derivative when the
spinor parameters are not necessarily parallel.  Instead we will take
them to be Killing: $\nabla_i \epsilon_L = \lambda_L \sigma_i
\epsilon_L$ and $\nabla_i \epsilon_R^\dagger = \lambda_R
\epsilon_R^\dagger \sigma_i$ for some (either real or imaginary)
constants $\lambda_L$ and $\lambda_R$.  We add terms
\begin{equation}
  \eL^{(3)} \rightsquigarrow \eL^{(3)} + \alpha_1 \Tr \chi^\dagger \psi + \half \alpha_2 \Tr \phi^2 + \alpha_3 \Tr \phi D + \half \alpha_4 \Tr D^2~
\end{equation}
to the lagrangian and also
\begin{equation}
  \begin{aligned}[m]
    \delta_L \psi & \rightsquigarrow \delta_L \psi + \beta_1 \phi \epsilon_L\\
    \delta_L D & \rightsquigarrow \delta_L D + \beta_2 \chi^\dagger \epsilon_L\\
  \end{aligned}
  \qquad\qquad
  \begin{aligned}[m]
    \delta_R \chi^\dagger & \rightsquigarrow \delta_R \chi^\dagger - \beta_3 \epsilon_R^\dagger \phi \\
    \delta_R D & \rightsquigarrow \delta_R D + \beta_4 \epsilon_R^\dagger \psi~,
  \end{aligned}
\end{equation}
for some constants $\alpha_1,\alpha_2,\alpha_3,\alpha_4,\beta_1,\beta_2,\beta_3,\beta_4$ to be determined.

We start by computing $\delta_L \eL^{(3)}$.  Using
equation~\eqref{eq:dlesymr3aux}, we arrive at (henceforth dropping
$\Tr$ from the notation)
\begin{multline}
  \delta_L \eL^{(3)} = \nabla_i X_L^i - \lambda_L (\half \varepsilon_{jk\ell}F^{jk} - D_\ell\phi) \chi^\dagger \sigma^\ell \epsilon_L - i \beta_1 \chi^\dagger \slashed{D} (\phi \epsilon_L) - \beta_2 D \chi^\dagger \epsilon_L\\
  + \alpha_1 \chi^\dagger \left((D+\beta_1 \phi) \epsilon_L + i(\half \varepsilon_{ijk}F^{ij}-D_k\phi)\sigma^k\epsilon_L \right) + \alpha_2 \phi \chi^\dagger \epsilon_L + \alpha_3 D \chi^\dagger \epsilon_L\\
  + \alpha_3 \phi \left(i\chi^\dagger \overleftarrow{\slashed{D}} \epsilon_L + \beta_2 \chi^\dagger \epsilon_L\right) + \alpha_4 D\left(i \chi^\dagger \overleftarrow{\slashed{D}} \epsilon_L + [\phi,\chi^\dagger \epsilon_L] + \beta_2 \chi^\dagger \epsilon_L\right)~,
\end{multline}
where $X_L^i = -i \chi^\dagger \left( \sigma^i D + \sigma_j F^{ij} -i D^i\phi \right)\epsilon_L $, and where we have used that $\sigma^i \sigma_j \sigma_i = - \sigma_j$.

The $\chi^\dagger F$ terms vanish provided that $\alpha_1 =
-i\lambda_L$, which also takes care of the $\chi^\dagger D_i\phi$
terms.  The $\chi^\dagger D A_i$ terms impose $\alpha_4 =0$, whereas
the $\chi^\dagger \phi A_i$ terms become a total derivative $\nabla_i
Y_L^i$, with $Y_L^i = -i \beta_1 \phi \chi^\dagger
\sigma^i\epsilon_L$, provided that $\alpha_3 = - \beta_1$.  The
$\chi^\dagger D$ terms vanish if $\beta_2 = -(\beta_1 + i \lambda_L)$
and the $\chi^\dagger \phi$ terms vanish provided that $\alpha_2 = -
\beta_1^2$.

In summary,
\begin{equation}
  \eL^{(3)} := -i \chi^\dagger \slashed{D} \psi - \chi^\dagger [\phi,\psi] - i\lambda_L \chi^\dagger \psi - \tfrac14 F^2 - \half |D\phi|^2 - \tfrac12 (D+\beta_1 \phi)^2
\end{equation}
transforms as
\begin{equation}
  \delta_L \eL^{(3)} = \nabla_i \left( -i \chi^\dagger \left( \sigma^i (D + \beta_1 \phi) + \sigma_j F^{ij} -i D^i\phi \right)\epsilon_L \right)~,
\end{equation}
under
\begin{equation}
  \begin{split}
    \delta_L A_i &= i \chi^\dagger \sigma_i \epsilon_L\\
    \delta_L \phi &= \chi^\dagger \epsilon_L\\
    \delta_L \chi^\dagger &= 0\\
    \delta_L \psi &= (D + \beta_1 \phi) \epsilon_L + \tfrac{i}2 \varepsilon_{ijk} F^{ij}\sigma^k \epsilon_L - i D_i\phi\sigma^i \epsilon_L \\
    \delta_L D &= i \chi^\dagger \overleftarrow{\slashed{D}} \epsilon_L + [\phi,\chi^\dagger\epsilon_L] - (\beta_1 + i \lambda_L) \chi^\dagger \epsilon_L~,
  \end{split}
\end{equation}
with $\nabla_i \epsilon_L = \lambda_L \sigma_i\epsilon_L$.

Notice that the action depends on $\lambda_L$, hence once the action is fixed, the sign of the Killing constant in the Killing spinor equation is also fixed.

Next we compute $\delta_R \eL^{(3)}$ and use equation~\eqref{eq:dresymr3aux} to find
\begin{multline}
  \delta_R \eL^{(3)} =  \nabla_i X^i_R - \lambda_R (\half \varepsilon_{jk\ell} + D_\ell\phi)\epsilon_R^\dagger \sigma^\ell \psi + i\beta_3 \phi \epsilon_R^\dagger \slashed{D} \psi - \beta_4 D \epsilon_R^\dagger \psi + \beta_1^2 \phi \epsilon_R^\dagger \psi\\
  + i \lambda_L \left((D+\beta_3\phi)\epsilon_R^\dagger \psi + i (\half \varepsilon_{ijk}F^{ij} + D_k \phi) \epsilon_R^\dagger \sigma^k \psi \right)\\
  + \beta_1 D \epsilon_R^\dagger \psi - \beta_1 \phi  (i\epsilon_R^\dagger \slashed{D}\psi + \beta_4 \epsilon_R^\dagger\psi)~,
\end{multline}
where we have again used $\sigma^i \sigma_j \sigma_i = - \sigma_j$ and where $X_R^i = \varepsilon^{ijk} \epsilon_R^\dagger \left(-\half F_{jk} +i D_j\phi \sigma_k\right) \psi$.

The $F\psi$ terms vanish provided that $\lambda_R = - \lambda_L$, and
this also takes care of the $D_i\phi \psi$ terms.  Notice that this
means that the vector field $\xi^i = 2i \epsilon_R^\dagger \sigma^i
\epsilon_L$ is a Killing vector, and not merely conformal Killing.
Indeed,
\begin{equation}
  \begin{split}
    \nabla_i \xi_j &= 2i \lambda_R \epsilon_R^\dagger \sigma_i \sigma_j \epsilon_L + 2i \lambda_L \epsilon_R^\dagger \sigma_j \sigma_i \epsilon_L\\
    &= -2i \lambda_L \epsilon_R^\dagger (\sigma_i \sigma_j - \sigma_j \sigma_i) \epsilon_L \\
    &= -2i \lambda_L \varepsilon_{ijk} \xi^k~,
  \end{split}
\end{equation}
whence $\nabla_i \xi_j + \nabla_j \xi_i = 0$.

The $A_i \phi \psi$ terms vanish provided that $\beta_3 = \beta_1$,
whereas the vanishing of the $D\psi$ terms set $\beta_4 = \beta_1 + i
\lambda_L$, which also takes care of the $\phi\psi$ terms.

In summary, and letting $\lambda_L = - \lambda_R = \lambda$,
\begin{equation}
  \eL^{(3)} := -i \chi^\dagger \slashed{D} \psi - \chi^\dagger [\phi,\psi] - i\lambda \chi^\dagger \psi - \tfrac14 F^2 - \half |D\phi|^2 - \tfrac12 (D+\beta_1 \phi)^2
\end{equation}
transforms as
\begin{equation}
  \delta_R \eL^{(3)} = \nabla_i \left(\varepsilon^{ijk} \epsilon_R^\dagger \left(-\half F_{jk} +i D_j\phi \sigma_k\right) \psi \right)~,
\end{equation}
under
\begin{equation}
  \begin{split}
    \delta_R A_i &= - i \epsilon_R^\dagger \sigma_i \psi\\
    \delta_R \phi &= - \epsilon_R^\dagger \psi\\
    \delta_R \chi^\dagger &= - (D+\beta_1 \phi) \epsilon_R^\dagger - i (\half \varepsilon_{ijk}F^{ij} + D_k\phi)\epsilon_R^\dagger \sigma^k\\
    \delta_R \psi &= 0\\
    \delta_R D &= i\epsilon_R^\dagger \slashed{D} \psi + \epsilon_R^\dagger [\phi,\psi] + (\beta_1 + i \lambda) \epsilon_R^\dagger \psi~,
  \end{split}
\end{equation}
with $\nabla_i \epsilon_L = \lambda \sigma_i\epsilon_L$ and $\nabla_i \epsilon_R^\dagger = - \lambda \epsilon_R^\dagger \sigma_i$.

One can show that the supersymmetry algebra closes as follows:
\begin{equation}
  [\delta_L,\delta'_L] = 0 = [\delta_R,\delta'_R] \qquad\text{and}\qquad [\delta_L,\delta_R] = \eL_\xi + \delta^{\text{gauge}}_\Lambda + \delta^R_\varpi~,
\end{equation}
for $\xi^i = 2 i \epsilon_R^\dagger \sigma^i \epsilon_L$ and $\Lambda = \xi^i A_i + 2 \epsilon_R^\dagger \epsilon_L \phi$, and where $\delta^R_\varpi$ is an R-symmetry transformation with $\varpi = -4\lambda \epsilon_R^\dagger \epsilon_L$, where
\begin{equation}
  \delta^R_\varpi \psi = i \varpi \psi \qquad\text{and}\qquad  \delta^R_\varpi \chi^\dagger = - i \varpi \chi^\dagger~.
\end{equation}
Indeed, it's induced from four-dimensions, where it is generated by
$\gamma^5$.  Notice that $\varpi$ is actually constant, so that this
is indeed a rigid R-symmetry transformation.  Similarly, it is worth
remarking that $\eL_\xi$ now means the spinorial Lie derivative
\cite{Kosmann1972} on the spinor fields, which in our case becomes
\begin{equation}
  \eL_\xi \psi = \xi^i \nabla_i \psi + \lambda \xi^i \sigma_i \psi \qquad\text{and}\qquad
  \eL_\xi \chi^\dagger = \xi^i \nabla_i \chi^\dagger - \lambda \xi^i \chi^\dagger \sigma_i~.
\end{equation}
One can check that this is indeed the expression which follows by
evaluating the definition $\eL_\xi = \nabla_\xi + \rho(A_\xi)$, with
$A_\xi$ the skew-symmetric endomorphism of the tangent bundle defined
by $A_\xi (X) = -\nabla_X \xi$ and where $\rho$ is the spin
representation.

The parameter $\beta_1$ remains free and can be set to zero if so
desired.  This is equivalent to the field redefinition $D
\rightsquigarrow D + \beta_1 \phi$.  Doing so, we have that the action
with lagrangian
\begin{equation}
  \label{eq:3dlagrangian}
  \eL^{(3)} = -i \chi^\dagger \slashed{D} \psi - \chi^\dagger [\phi,\psi] - i\lambda \chi^\dagger \psi - \tfrac14 F^2 - \half |D\phi|^2 - \tfrac12 D^2
\end{equation}
transforms as
\begin{align}
  \delta_L \eL^{(3)} &= \nabla_i \left( -i \chi^\dagger \left( \sigma^i D + \sigma_j F^{ij} -i D^i\phi \right)\epsilon_L \right)\\
  \delta_R \eL^{(3)} &= \nabla_i \left(\varepsilon^{ijk} \epsilon_R^\dagger \left(-\half F_{jk} +i D_j\phi \sigma_k\right) \psi \right)
\end{align}
under
\begin{equation}
  \label{eq:susytrans}
  \begin{aligned}[m]
    \delta_L A_i &= i \chi^\dagger \sigma_i \epsilon_L\\
    \delta_L \phi &= \chi^\dagger \epsilon_L\\
    \delta_L \chi^\dagger &= 0\\
    \delta_L \psi &= D \epsilon_L + i (\half \varepsilon_{ijk} F^{ij} - D_k\phi)\sigma^k\epsilon_L \\
    \delta_L D &= i \chi^\dagger \overleftarrow{\slashed{D}} \epsilon_L + [\phi,\chi^\dagger]\epsilon_L - i \lambda \chi^\dagger \epsilon_L~,
  \end{aligned}
  \qquad\qquad
  \begin{aligned}[m]
    \delta_R A_i &= - i \epsilon_R^\dagger \sigma_i \psi\\
    \delta_R \phi &= - \epsilon_R^\dagger \psi\\
    \delta_R \chi^\dagger &= - D \epsilon_R^\dagger - i (\half \varepsilon_{ijk}F^{ij} + D_k\phi)\epsilon_R^\dagger \sigma^k\\
    \delta_R \psi &= 0\\
    \delta_R D &= i\epsilon_R^\dagger \slashed{D} \psi + \epsilon_R^\dagger [\phi,\psi] + i \lambda \epsilon_R^\dagger \psi~,
  \end{aligned}
 \end{equation}
with $\nabla_i \epsilon_L = \lambda \sigma_i\epsilon_L$ and $\nabla_i \epsilon_R^\dagger = - \lambda \epsilon_R^\dagger \sigma_i$.

\subsection{Some remarks}
\label{sec:some-remarks}

The first remark is that there is only a mass term for the fermions,
yet none for the scalar.  (This is a choice.)  The choice of $\lambda$
is dictated by the geometry up to a sign, but that sign is immaterial
since $\lambda$ appears in the action.

Secondly, it seems that the action is not ``exact'' in that $\eL^{(3)}
\epsilon_R^\dagger \epsilon_L \neq \delta_L\delta_R \Xi$ for any
reasonable
$\Xi$.

Thirdly, we remark that this theory agrees morally with one of the
theories in Family A in \cite{Blau2000}. In fact, if we eliminate the
auxiliary field, then it agrees with the theory described by equation
(3.10) in that paper, denoted $N=2$ in $d=3$.

Finally, let us comment on the geometry of the manifolds admitting
Killing spinors.  The integrability condition for solutions of the
Killing spinor equation $\nabla_i \epsilon_L = \lambda
\sigma_i\epsilon_L$ says that the metric is Einstein.  The vanishing
of the Weyl tensor in three dimensions implies that the Riemann
curvature tensor of a Einstein three-dimensional riemannian manifold
can be written purely in terms of the scalar curvature and the metric;
in other words, it has constant sectional curvature, where the value
of the scalar curvature is related to the Killing constant $\lambda$
by $R = -24 \lambda^2$ in our conventions.  Therefore the existence of
Killing spinors with real $\lambda$ forces the manifold to be
hyperbolic, whereas for imaginary $\lambda$ it would be spherical.  In
the simply-connected case, we have three-dimensional hyperbolic space
and the 3-sphere, respectively, which admit the maximum number of such
Killing spinors, with either sign of the Killing constant.

\section{Moduli space of BPS configurations}
\label{sec:moduli-space-bps}

In this section we start the analysis of the geometry of the moduli
space of BPS configurations.  The first observation, which is crucial
for this approach to the problem, is that the BPS configurations are
precisely the BPS monopoles with $D=0$.  More precisely, bosonic
configurations for which $\delta_L\psi=0$ are precisely those obeying
$D=0$ and $D_k\phi = \half \varepsilon_{ijk} F^{ij}$, for which the
$\delta_L$ supersymmetries with parameter $\epsilon_L$ obeying
$\nabla_i \epsilon_L = \lambda \sigma_i \epsilon_L$ are preserved.
This is easy to see by writing
\begin{equation}
  \delta_L\psi = (D + i (\half \varepsilon_{ijk} F^{ij} -
  D_k\phi)\sigma^k)\epsilon_L
\end{equation}
and noticing that the determinant of $D + i (\half \varepsilon_{ijk}
F^{ij} - D_k\phi)\sigma^k$ is zero if and only if $D=0$ and $\half
\varepsilon_{ijk} F^{ij} - D_k\phi=0$.  Similarly, the bosonic
configurations with $D_k\phi = - \half \varepsilon_{ijk} F^{ij}$ and
$D=0$ are precisely the ones which preserve the $\delta_R$
supersymmetries with parameter $\epsilon_R^\dagger$ obeying $\nabla_i
\epsilon_R^\dagger = -\lambda \epsilon_R^\dagger \sigma_i$.  It is the
these latter bosonic BPS configurations whose moduli space $\eM$ we
will study in the rest of this paper.  The moduli space $\eM$ is
defined as the quotient $\eP/\eG$ of the space $\eP$ of solutions of
the Bogomol'nyi equation
\begin{equation}
  \label{eq:bogeqn}
  D_i\phi + \varepsilon_{ijk} F^{jk} = 0
\end{equation}
by the action of the group $\eG$ of gauge transformations:
\begin{equation}
  \label{eq:gaugetrans}
  A \mapsto g A g^{-1} - dg g^{-1} \qquad\text{and}\qquad \phi \mapsto
  g \phi g^{-1}~,
\end{equation}
where $g: H^3 \to G$ is a smooth function.  We mention once again that
the euclidean theory has complex fields, so that strictly speaking the
half-BPS states actually correspond to complexified hyperbolic
monopoles with $D=0$.

\subsection{Zero modes}
\label{sec:zero-modes}

Consider a one-parameter family $A_i(s), \phi(s)$ of bosonic BPS
configurations, where $s$ is a formal parameter.  This means that for
all $s$, $A_i(s)$ and $\phi(s)$ obey the Bogomol'nyi
equation \begin{equation}
  \label{eq:bogeqnt}
  D_i(s)\phi(s) + \varepsilon_{ijk} F^{jk}(s) = 0~.
\end{equation}
Differentiating with respect to $s$ at $s=0$, we find
\begin{equation}
  \label{eq:linbogeqn}
  D_i(0) \dot\phi - [\phi(0),\dot A_i] + \varepsilon_{ijk} D^j(0) \dot A^k = 0~,
\end{equation}
where $\dot A_i = \left.\frac{\partial A_i}{\partial s}\right|_{s=0}$,
$\dot \phi = \left.\frac{\partial \phi}{\partial s}\right|_{s=0}$ and
$D_i(0) = \partial_i + [A_i(0),-]$.  Equation~\eqref{eq:linbogeqn} is
the linearisation at $(A_i(0),\phi(0))$ of the Bogomol'nyi equation
and solutions of that equation will be termed \emph{bosonic zero
  modes}.

One way to generate bosonic zero modes is to consider the tangent
vector to the orbit of a one-parameter subgroup of the group of gauge
transformations.  The subspace of such zero modes is the tangent space
to the gauge orbit of $(A_i(0),\phi(0))$.  The true tangent space to
the moduli space can be identified with a suitable complement of that
subspace.  A choice of such a complement is essentially a choice of
connection on the principal $\eG$-bundle $\eP \to \eM$.  In the
absence of a natural riemannian metric on $\eP$, we will employ
supersymmetry to define this connection.

Supersymmetry relates the bosonic zero modes to \emph{fermionic zero
  modes} $\dot\psi$ which are solutions of the (already linear) field
equations for $\psi$ at $(A_i(0),\phi(0))$:
\begin{equation}
  \label{eq:fer0modes}
  \slashed{D}(0)\dot\psi - i [\phi(0),\dot\psi] + \lambda \dot\psi = 0~.
\end{equation}

Let $\eta,\zeta$ be Killing spinors on hyperbolic space satisfying
\begin{equation}
  \nabla_i \eta = \lambda \sigma_i \eta \qquad\text{and}\qquad \nabla_i \zeta^\dagger = - \lambda \zeta^\dagger \sigma_i~.
\end{equation}
Of course, hyperbolic space has the maximal number of either class of
such Killing spinors.

Let $(\dot A_i,\dot\phi)$ satisfy the linearised Bogomol'nyi
equation~\eqref{eq:linbogeqn} and let
\begin{equation}
  \label{eq:b2fzm}
  \dot\psi = i \dot A_i \sigma^i \eta - \dot\phi \eta~.
\end{equation}
We claim that $\dot\psi$ so defined is a fermionic zero mode provided
that $(\dot A_i,\dot\phi)$ obey in addition the \emph{generalised
  Gauss law}
\begin{equation}
  \label{eq:gausslaw}
  D^i(0) \dot A_i + [\phi(0),\dot\phi] + 4 i \lambda \dot\phi = 0~.
\end{equation}
Indeed, with the tacit evaluation at $s=0$,
\begin{align*}
  &\slashed{D}\left(i \dot A_i \sigma^i \eta - \dot\phi \eta\right) + i \left[ \left(i \dot A_i \sigma^i \eta - \dot\phi \eta\right),\phi \right] + \lambda \left(i \dot A_i \sigma^i \eta - \dot\phi \eta\right)\\
  &= i D_j \dot A_i \sigma^j\sigma^i \eta + i \dot A_i\sigma^j\sigma^i \nabla_j \eta - D_i \dot\phi \sigma^i\eta - \dot\phi\slashed{\nabla}\eta - [\dot A_i,\phi] \sigma^i \eta - i [\dot\phi,\phi]\eta + i \lambda\dot A_i \sigma^i \eta - \lambda \dot\phi \eta\\
  &= i D^i \dot A_i \eta - \varepsilon^{ijk} D_i \dot A_j \sigma_k \eta 
- D_i \dot\phi \sigma^i\eta - 4\lambda \dot \phi \eta
 - [\dot A_i,\phi] \sigma^i \eta - i [\dot\phi,\phi]\eta~,
\end{align*}
where we have used that $\sigma^j \sigma_i \sigma_j = - \sigma_i$ and
that $\slashed{\nabla} \eta = 3 \lambda \eta$.  We can rewrite the
resulting expression as follows
\begin{equation}
  \left(i D^i \dot A_i - i [\dot\phi,\phi] - 4\lambda \dot \phi\right) \eta -
\left( \varepsilon^{ijk} D_i \dot A_j + D^k \dot\phi + [\dot A^k,\phi] \right) \sigma_k \eta~,
\end{equation}
which contains two kinds of terms: those which are proportional to
$\sigma_k \eta$ vanish because of the linearised Bogomol'nyi
equation~\eqref{eq:linbogeqn}, whereas the ones proportional to $\eta$
cancel if and only if the generalised Gauss law~\eqref{eq:gausslaw} is
satisfied.

One might be surprised by the last term in the generalised Gauss law
as this is absent in the case of euclidean monopoles.  And indeed, we
see that in the flat space limit $\lambda \to 0$ this term disappears.
The Gauss law is a gauge-fixing condition, or more geometrically, it
is an Ehresmann connection on the principal gauge bundle $\eP \to \eM$
over the moduli space; that is, a $\eG$-invariant complement to the
tangent space to the gauge orbit through every point of $\eP$.  It is
not hard to see that condition~\eqref{eq:gausslaw} is $\eG$-invariant
and that it provides a complement to the gauge orbits.  However it is
not, as in the case of euclidean monopoles, the perpendicular
complement to the tangent space to the gauge orbits relative to a
$\eG$-invariant metric on $\eP$.

Conversely, if $\dot\psi$ obeys equation~\eqref{eq:fer0modes}, then
\begin{equation}
  \label{eq:f2bzm}
  \dot A_i = - i \zeta^\dagger \sigma_i \dot\psi \qquad\text{and}\qquad \dot\phi = - \zeta^\dagger \dot\psi
\end{equation}
obey the linearised Bogomol'nyi equation~\eqref{eq:linbogeqn} and the
generalised Gauss law~\eqref{eq:gausslaw}.  Indeed, and again with the
tacit evaluation at $s=0$,
\begin{align*}
  D_i \left(-\zeta^\dagger \dot \psi\right) + &\varepsilon_{ijk} D^j \left(-i\zeta^\dagger \sigma^k \dot \psi\right) - \left[\phi,\left(-i\zeta^\dagger \sigma_i \dot \psi\right)\right]\\
  &= -\nabla_i \zeta^\dagger \dot\psi - \zeta^\dagger D_i \dot\psi - i \varepsilon_{ijk} \nabla^j \zeta^\dagger \sigma^k \dot\psi - i \varepsilon_{ijk} \zeta^\dagger \sigma^k D^j\dot\psi + i \zeta^\dagger \sigma_i [\phi,\dot\psi]\\
  &= \lambda \zeta^\dagger \sigma_i \dot\psi - \zeta^\dagger D_i\dot\psi + i \zeta^\dagger [\phi,\dot\psi] + i \lambda \varepsilon_{ijk} \zeta^\dagger \sigma^{jk}\dot\psi - i \varepsilon_{ijk} \zeta^\dagger \sigma^k D^j \dot\psi~.
\end{align*}
We now use that $\varepsilon_{ijk}\sigma^{jk} = 2 i \sigma_i$ and that $i[\phi,\dot\psi]= \slashed{D}\dot\psi + \lambda \dot\psi$ to arrive at
\begin{align*}
  D_i \left(-\zeta^\dagger \dot \psi\right) + \varepsilon_{ijk} D^j \left(-i\zeta^\dagger \sigma^k \dot \psi\right) &- \left[\phi,\left(-i\zeta^\dagger \sigma_i \dot \psi\right)\right]\\
  &= - \zeta^\dagger D_i \dot\psi - i \varepsilon_{ijk} \zeta^\dagger \sigma^k D^j \dot\psi + \zeta^\dagger \sigma_i \slashed{D} \dot\psi~,
\end{align*}
which is seen to vanish after using that $\sigma_i\sigma_j = g_{ij} + i \varepsilon_{ijk} \sigma^k$ to expand $\sigma_i \slashed{D} \dot\psi$.

\subsection{A four-dimensional formalism}
\label{sec:four-dimens-form}

It is convenient for calculations to introduce a four-dimensional
language.  This amounts to working on the four-dimensional manifold
$H^3 \times S^1$, but where the fields are invariant under
translations in $S^1$.  The relevant Clifford algebra is now generated
by $\Gamma_\mu = (\Gamma_i,\Gamma_4)$ given by
\begin{equation}
  \Gamma_i =
  \begin{pmatrix}
    0 & \sigma_i \\ \sigma_i & 0
  \end{pmatrix}
  \qquad
  \Gamma_4 =
  \begin{pmatrix}
    0 & i \\ -i & 0
  \end{pmatrix}
\end{equation}
which satisfy $\Gamma_\mu \Gamma_\nu + \Gamma_\nu \Gamma_\mu = 2 \delta_{\mu\nu} \II$.  Let $\zeta_R = \begin{pmatrix}0 \\ \zeta \end{pmatrix}$ and
$\eta_R = \begin{pmatrix}0 \\ \eta \end{pmatrix}$, which obey the Killing spinor equations
\begin{equation}
\label{eq:derkill}
  \nabla_i \eta_R = -i \lambda \Gamma_i \Gamma_4 \eta_R \qquad\text{and}\qquad
  \nabla_i \zeta_R^\dagger = -i \lambda \zeta_R^\dagger \Gamma_4 \Gamma_i~,
\end{equation}
and in addition $\nabla_4 \eta_R = 0$ and $\nabla_4 \zeta_R^\dagger = 0$.  The zero modes are now $\dot\Psi_L = \begin{pmatrix}\dot\psi \\ 0 \end{pmatrix}$ and $\dot A_\mu = (\dot A_i, \dot \phi)$ and the relations \eqref{eq:b2fzm} and \eqref{eq:f2bzm} between them can now be rewritten respectively as
\begin{equation}
  \dot \Psi_L = i \dot A_\mu \Gamma^\mu \eta_R \qquad\text{and}\qquad
  \dot A_\mu = - i \zeta_R^\dagger \Gamma_\mu \dot\Psi_L~.
\end{equation}

It is perhaps pertinent to remark that these equations are not meant
to be understood as mutual inverse relations; that is, substituting
the first equation for $\dot A_\mu$ in the second equation does
\emph{not} lead to an identity and neither does substituting the
second equation for $\dot \Psi_L$ into the first.  What these
relations do mean is that given a bosonic zero mode $\dot A_\mu$ and a
Killing spinor $\eta$ on $H^3$, the RHS of the second of the above
equations defines a fermionic zero mode; and that, conversely, given a
fermionic zero mode $\dot \Psi_L$ and a Killing spinor $\zeta$ on
$H^3$, the RHS of the first of the above equations defines a bosonic
zero mode.

More formally, let us define the vector spaces
\begin{equation}
\label{eq:killset}
  K^\pm = \left\{ \xi_R \middle | \nabla_i \xi_R = \mp i \lambda
  \Gamma_i \Gamma_4 \xi_R \quad\text{and}\quad \nabla_4 \xi_R =
  0\right\}~.
\end{equation}
$K^\pm$ is a two-dimensional complex vector space isomorphic to the
vector space of Killing spinor fields on $H^3$ with the stated sign of
the Killing constant; that is,
\begin{equation}
  K^\pm \cong \left\{ \xi \middle | \nabla_i \xi = \pm \lambda \sigma_i \xi\right\}~.
\end{equation}
Then letting $Z_0$ and $Z_1$ stand for the vector spaces of
(complexified) bosonic and fermionic zero modes, respectively, we have
exhibited real bilinear maps
\begin{equation}
  \label{eq:Z0andZ1}
  \begin{aligned}[m]
    K^+ \times Z_0 &\to Z_1\\
    (\eta_R,\dot A_\mu) &\mapsto i \dot A_\mu \Gamma^\mu \eta_R
  \end{aligned}
  \qquad\text{and}\qquad
  \begin{aligned}[m]
    K^- \times Z_1 &\to Z_0\\
    (\zeta_R,\dot \Psi_L) &\mapsto -i \zeta_R^\dagger \Gamma_\mu \dot\Psi_L~.
  \end{aligned}
\end{equation}
We may compose the maps to arrive at
\begin{equation}
  \label{eq:Z0toZ0} 
  \begin{aligned}[m]
    K^+ \times K^- \times Z_0 &\to Z_0\\
    (\eta_R,\zeta_R, \dot A_\mu) &\mapsto \zeta_R^\dagger \eta_R \dot
    A_\mu + \zeta_R^\dagger \Gamma_\mu{}^\nu \eta_R \dot A_\nu
  \end{aligned}
\end{equation}
and
\begin{equation}
  \label{eq:Z1toZ1}
  \begin{aligned}[m]
    K^+ \times K^- \times Z_1 &\to Z_1\\
    (\eta_R,\zeta_R, \dot \Psi_L) &\mapsto 2 \zeta_R^\dagger \eta_R \dot \Psi_L~,
  \end{aligned}
\end{equation}
where in deriving these identities we have used the Fierz
identity~\eqref{eq:fierzRR} for commuting spinors.

If we fix $\zeta_R$ and $\eta_R$ such that $\zeta_R^\dagger \eta_R =
\tfrac12$, which we can always do, then the composite map in
equation~\eqref{eq:Z1toZ1} is the identity, which implies that the
maps in equation~\eqref{eq:Z0andZ1} are invertible. In particular,
this implies that the vector spaces $Z_0$ and $Z_1$ of (complexified)
bosonic and fermionic zero modes, respectively, are isomorphic. This
is the hyperbolic analogue of the result of Zumino \cite{Zumino} for
euclidean monopoles. That result can be rederived without using
supersymmetry via the calculation of the index of the Dirac operator
in the presence of a monopole. For hyperbolic monopoles this
calculation has not been performed, to our knowledge, but it is
conceivable that it may be possible using the generalisation of the
Callias index theorem \cite{Callias} in \cite{MR1266069}.

We end this section by recording that in four-dimensional language the
fermionic zero modes are defined by the equation
\begin{equation}
  \slashed{D} \dot\Psi_L = -i \lambda \Gamma_4 \dot\Psi_L~,
\end{equation}
whereas those defining the bosonic zero modes are
\begin{equation}
  D_{[\mu} \dot A_{\nu]} = - \half \varepsilon_{\mu\nu\rho\sigma} D^\rho \dot A^\sigma \qquad\text{and}\qquad
  D^\mu \dot A_\mu = - 4 i \lambda \dot A_4~.
\end{equation}
The first equation is simply the statement that the $\fg$-valued 2-form $D_{[\mu} \dot A_{\nu]}$ is antiselfdual.

\subsection{Complex structures}
\label{sec:complex-structures}

We start by defining some natural endomorphisms of the complexified
tangent bundle of $H^3 \times S^1$ which can be built out of the
Killing spinors.

Let us choose a complex basis $\eta_{R\alpha}$ and $\zeta_{R\beta}$,
for $\alpha,\beta=1,2$, for the vector spaces $K^+$ and $K^-$ of
Killing spinors, respectively, which satisfies in addition the
normalisation condition $\zeta_{R\alpha}^\dagger \eta_{R\beta} =
\delta_{\alpha\beta}$.  Let $A_{\alpha\beta}$ be the endomorphism of
$T_\CC(H^3 \times S^1)$ defined by
\begin{equation}
  A_{\alpha\beta\,\mu}{}^\nu = -i \zeta_{R\alpha}^\dagger \Gamma_\mu{}^\nu \eta_{R\beta}~.
\end{equation}
Then one can show that the linear combinations
\begin{equation}
  I = A_{11} \qquad J = \half (A_{12} + A_{21}) \qquad K =-\frac{i}2 (A_{12}-A_{21})
\end{equation}
satisfy the quaternion algebra
\begin{equation}
  I^2 = J^2 = - \II \qquad I J = - J I = K ~.
\end{equation}

More invariantly, if $\eta_R \in K^+$ and $\zeta_R \in K^-$, let
\begin{equation}
  E_\mu{}^\nu = -i \zeta_R^\dagger \Gamma_\mu{}^\nu \eta_R
\end{equation}
denote the corresponding endomorphism of $T_\CC(H^3 \times S^1)$.  It
follows from the fact that $\eta_R,\zeta_R$ have negative chirality,
i.e., $\Gamma_{1234}\eta_R = - \eta_R$ and similarly for $\zeta_R$,
that $E_{\mu\nu}$ is self-dual:
\begin{equation}
  \label{eq:E_is_SD}
  \tfrac12 \varepsilon_{\mu\nu\rho\sigma} E^{\rho\sigma} = E_{\mu\nu}~,
\end{equation}
and also that
\begin{equation}
  \label{eq:CplxStruct}
  E_\mu{}^\rho E_\rho{}^\nu = - (\zeta_R^\dagger \eta_R)^2 \delta_\mu{}^\nu~.
\end{equation}
The proof of this expression follows from the Fierz
identity~\eqref{eq:fierzRR} and tedious use of the Clifford relations.
Hence if we choose $\eta_R$ and $\zeta_R$ such that $\zeta_R^\dagger
\eta_R = 1$, then the endomorphism $E$ is a (complex-linear) almost
complex structure on $T_\CC(H^3 \times S^1)$.

In addition, from from the fact that $\eta_R,\zeta_R$ are Killing spinors it also follows that
\begin{equation}
  \label{eq:CovDerE}
  \nabla_4 E_{\mu\nu} = 0, \qquad \nabla_i E_{4j} = 2 i \lambda E_{ij}
  \qquad\text{}\qquad \nabla_i E_{jk} = -2 i \lambda \left(\delta_{ij}E_{4k} - \delta_{ik} E_{4j}\right)~.
\end{equation}

Indeed, the first equation follows from the fact that $\nabla_4
\zeta_R = 0 = \nabla_4 \eta_R$.  The second equation follows from the
following calculation:
\begin{equation}
  \begin{split}
    \nabla_i E_{4j} &= \nabla_i \left(-i \zeta_R^\dagger \Gamma_4 \Gamma_j \eta_R\right)\\
    &= -i \left(-i \lambda \zeta_R^\dagger \Gamma_4 \Gamma_i\right)\Gamma_4\Gamma_j \eta_R - i \zeta_R^\dagger \Gamma_4 \Gamma_j \left(-i \lambda \Gamma_i \Gamma_4 \eta_R\right)\\
    &= - \lambda \zeta_R^\dagger \Gamma_4 \Gamma_i \Gamma_4\Gamma_j \eta_R -
  \lambda \zeta_R^\dagger \Gamma_4 \Gamma_j \Gamma_i \Gamma_4 \eta_R\\
    &= \lambda \zeta_R^\dagger \left(\Gamma_i \Gamma_j - \Gamma_j \Gamma_i\right) \eta_R\\
    &= 2\lambda \zeta_R^\dagger \Gamma_{ij} \eta_R\\
    &= 2i \lambda E_{ij}~,
  \end{split}
\end{equation}
where we have used the Clifford relations and the fact that $\nabla_i
\zeta_R^\dagger = - i \lambda \zeta_R^\dagger \Gamma_4 \Gamma_i$.

The third and final equation follows from a similar calculation:
\begin{equation}
  \begin{split}
    \nabla_i E_{jk} &= \nabla_i \left(-i \zeta_R^\dagger \Gamma_{jk} \eta_R \right)\\
    &= -i \left(-i \lambda \zeta_R^\dagger \Gamma_4 \Gamma_i\right)\Gamma_{jk} \eta_R - i \zeta_R^\dagger \Gamma_{jk} \left(-i \lambda \Gamma_i \Gamma_4 \eta_R\right)\\
    &= -\lambda \zeta_R^\dagger \Gamma_4 \Gamma_i \Gamma_{jk} \eta_R - \lambda \zeta_R^\dagger \Gamma_{jk}\Gamma_i \Gamma_4 \eta_R\\
    &= -\lambda \zeta_R^\dagger \Gamma_4 \left(\Gamma_i \Gamma_{jk} - \Gamma_{jk} \Gamma_i\right) \eta_R~.
  \end{split}
\end{equation}
We now use the following consequences of the Clifford relations:
\begin{equation}
  \Gamma_i \Gamma_{jk} = \Gamma_{ijk} + \delta_{ij} \Gamma_k - \delta_{ik} \Gamma_j \qquad\text{and}\qquad \Gamma_{jk} \Gamma_i = \Gamma_{jki} + \delta_{ik} \Gamma_j - \delta_{ij} \Gamma_k
\end{equation}
whence
\begin{equation}
  \Gamma_i \Gamma_{jk} - \Gamma_{jk} \Gamma_i = 2 \delta_{ij} \Gamma_k - 2 \delta_{ik} \Gamma_j~,
\end{equation}
and hence
\begin{equation}
  \begin{split}
    \nabla_i E_{jk} &= -\lambda \zeta_R^\dagger \Gamma_4 \left(2 \delta_{ij} \Gamma_k - 2 \delta_{ik} \Gamma_j\right) \eta_R\\
    &= -2\lambda \delta_{ij} \zeta_R^\dagger \Gamma_4 \Gamma_k \eta_R + 2\lambda \delta_{ik} \zeta_R^\dagger \Gamma_4 \Gamma_j \eta_R\\
    &= -2 i \lambda \left(\delta_{ij} E_{4k} - \delta_{ik} E_{4j}\right)~.
  \end{split}
\end{equation}

Now we show that the endomorphisms $E_\mu{}^\nu$ act naturally on the
bosonic zero modes $\dot A_\mu$.  In other words, we show that if
$\dot A_\mu$ obeys the linearised Bogomol'nyi
equation~\eqref{eq:linbogeqn} and the generalised Gauss
law~\eqref{eq:gausslaw}, then so does its image $\dot B_\mu :=
E_\mu{}^\nu \dot A_\nu$ under such an endomorphism.

We start with the generalised Gauss law~\eqref{eq:gausslaw}.  By definition,
\begin{equation}
  \begin{split}
    D^\mu \dot B_\mu &= D^\mu \left(E_\mu{}^\nu \dot A_\nu\right)\\
    &= \nabla^\mu E_\mu{}^\nu \dot A_\nu + E^{\mu\nu} D_\mu A_\nu\\
    &= \nabla^i E_i{}^\nu \dot A_\nu + E^{\mu\nu} D_{[\mu} A_{\nu]}\\
    &= -4i \lambda E_4{}^j \dot A_j\\
    &= -4i \lambda \dot B_4~,
  \end{split}
\end{equation}
where we have used equation~\eqref{eq:CovDerE} and the fact that,
since $E^{\mu\nu}$ is selfdual and $D_{[\mu} A_{\nu]}$ antiselfdual,
their inner product vanishes.  Thus we see that $\dot B_\mu$ obeys the
generalised Gauss law~\eqref{eq:gausslaw}.

Next we show that $\dot B_\mu$ obeys the linearised Bogomol'nyi
equation~\eqref{eq:linbogeqn}, which says that $D_{[\mu} \dot
B_{\nu]}$ is antiselfdual, or equivalently, that
\begin{equation}
  D_i \dot B_4 + \varepsilon_{ijk} D_j \dot B_k = 0~.
\end{equation}
Using equations~\eqref{eq:E_is_SD} and \eqref{eq:CovDerE}, we
calculate the first term in the left-hand side:
\begin{equation}
  \begin{split}
    D_i \dot B_4  &= 
    D_i \left(E_{4j} \dot A_j\right)\\
    &= \nabla_i E_{4j} \dot A_j + E_{4j} D_i \dot A_j\\
    &= 2 i \lambda E_{ij} \dot A_j + E_{4j} D_i \dot A_j\\
    &= - 2i \lambda \varepsilon_{ijk} E_{4k} \dot A_j + E_{4j} D_i \dot A_j~,
  \end{split}
\end{equation}
and then also the second term:
\begin{equation}
  \begin{split}
    \varepsilon_{ijk} D_j \dot B_k &= \varepsilon_{ijk} D_j \left(E_{kl} \dot A_l + E_{k4} \dot A_4 \right)\\
    &= \varepsilon_{ijk} \left(\nabla_j E_{kl} \dot A_l - \nabla_j E_{4k} \dot A_4 + E_{kl} D_j \dot A_l + E_{k4} D_j \dot A_4 \right)\\
    &= \varepsilon_{ijk} \left( 2 i \lambda E_{4k} \dot A_j - 2 i \lambda \varepsilon_{jkl} E_{4l} \dot A_4 + E_{kl} D_j \dot A_l - E_{4k} D_j \dot A_4 \right)\\
    &= 2 i \lambda \varepsilon_{ijk} E_{4k} \dot A_j + 4 i \lambda E_{4i} \dot A_4 - \varepsilon_{ijk} \varepsilon_{klm} E_{4m} D_j \dot A_l - E_{4k} \varepsilon_{ijk} D_j \dot A_4\\
    &= 2 i \lambda \varepsilon_{ijk} E_{4k} \dot A_j - E_{4j} D_j \dot A_i - E_{4k} \varepsilon_{ijk} D_j \dot A_4\\
    &= 2 i \lambda \varepsilon_{ijk} E_{4k} \dot A_j - E_{4j} D_j \dot A_i - E_{4k} (D_i \dot A_k - D_k \dot A_i)\\
    &= 2 i \lambda \varepsilon_{ijk} E_{4k} \dot A_j - E_{4k} D_i \dot A_k~,
  \end{split}
\end{equation}
where we have used that $\dot A_\mu$ obeys the linearised Bogomol'nyi equation~\eqref{eq:linbogeqn} and the generalised Gauss law~\eqref{eq:gausslaw}.  Finally, we notice that the sum of the two terms vanish.

In summary, we have shown that the vector $E_{\mu}{}^{\nu}\dot
A_{\nu}$ is tangent to the moduli space.  Since there is a quaternion
algebra in the span of the endomorphisms $E_\mu{}^\nu$, we see that
the complexified tangent space to the moduli space is a quaternionic
vector space.  Indeed, if we let $\dot A_{a\mu}$ denote a complex
frame for the complexified tangent space to $\eM$ at $(A,\phi)$, then
we may define endomorphisms $\eI$, $\eJ$ and $\eK$ of the tangent
space at that point by
\begin{equation}
\label{eq:endoontan}
  \eI_a{}^b \dot A_{b\mu} = I_\mu{}^\nu \dot A_{a\nu} \qquad
  \eJ_a{}^b \dot A_{b\mu} = J_\mu{}^\nu \dot A_{a\nu} \qquad
  \eK_a{}^b \dot A_{b\mu} = K_\mu{}^\nu \dot A_{a\nu}~.
\end{equation}
Letting the point $(A,\phi)$ vary we obtain a field of endomorphisms
of $T_\CC\eM$ which we also call $\eI,\eJ,\eK$.  It is evident that
just like $I,J,K$ generate a quaternion algebra, so do $\eI,\eJ,\eK$.

It is worth emphasising that $\eI,\eJ,\eK$ are \emph{complex linear}
endomorphisms of $T_\CC\eM$; that is, they commute with the complex
structure introduced when we complexified the tangent bundle of $\eM$.
That complex structure is unrelated to $\eI$, $\eJ$ and $\eK$.  In
fact, what we have is an action of the quaternions, say, on the right
and an action of the complex numbers on the left, whence an action of
$\CC\otimes_\RR\HH \cong \Mat(2,\CC)$.

\section{Geometry of the moduli space}
\label{sec:geom-moduli-space}

In order to probe the geometry of the moduli space $\eM$ of hyperbolic
monopoles, we will consider the multiplet corresponding to a
one-dimensional sigma model, except that we do not have an action for
this model. In other words, we consider maps $X : \RR \to \eM$,
$t\mapsto X(t)$, and the associated fermions $\theta$ which are
sections of $\Pi X^*T_\CC\eM$: the (oddified) pullback by $X$ of the
complexified tangent bundle of $\eM$. In this section we will first
linearise the supersymmetry transformations and in this way arrive at
an expression for the supersymmetry transformations of the bosonic
moduli. We will then derive the supersymmetry transformations of the
fermionic moduli by demanding closure of the one-dimensional $N=4$
supersymmetry algebra. This will also reveal the geometry of the
moduli space to be that of a pluricomplex manifold.

\subsection{Linearising the supersymmetry transformations}
\label{sec:Lin-susy-trans}

In this section we will derive the supersymmetry transformations for
the bosonic zero modes by linearising the supersymmetry
transformations preserved by the monopoles.

The $\delta_R$ supersymmetry transformations preserved by hyperbolic
monopole configurations are given by equation~\eqref{eq:susytrans}.
On the gauge field, and in four-dimensional language, it can be
written as
\begin{equation}
  \delta_\epsilon A_\mu = - i \epsilon_R^\dagger \Gamma_\mu
  \Psi_L~,
\end{equation}
which is already linear, hence at the level of the zero modes becomes
\begin{equation}
  \label{eq:linsusytrans}
  \delta_\epsilon \dot A_\mu = - i \epsilon_R^\dagger \Gamma_\mu
  \dot \Psi_L~.
\end{equation}
Choose a basis $\dot\Psi_{L a}$ for the space $Z_1$ of fermionic zero
modes.  This defines a basis $\dot A_{a\mu}$ for the space $Z_0$ of
complexified bosonic zero modes via the second map in
equation~\eqref{eq:Z0andZ1}: namely,
\begin{equation}
  \dot A_{a\mu} := -i \zeta_R^\dagger \Gamma_\mu \dot\Psi_{L a}~,
\end{equation}
where $\zeta_R \in K^-$ is a fixed Killing spinor.  From equation
\eqref{eq:Z1toZ1} we may invert this to write $\dot\Psi_{L a} = i\dot
A_{a\mu} \Gamma^\mu \eta_R$ for some $\eta_R\in K^+$ such that
$\zeta_R^\dagger \eta_R = \half$.

We now expand the general bosonic zero mode $\dot A_\mu = \dot
A_{a\mu} X^a$ as a linear combination of the basis $\dot A_{a\mu}$
and similarly for the general fermionic zero mode $\dot\Psi_L =
\dot\Psi_{L  a}\theta^a$.  Inserting this in
equation~\eqref{eq:linsusytrans}, we obtain
\begin{equation}
  \delta_\epsilon \dot A_\mu = \dot A_{a\mu} \delta_\epsilon X^a =
 \dot A_{a\nu} \epsilon_R^\dagger \Gamma_\mu \Gamma^\nu \eta_R
  \theta^a = 
  \dot A_{a\mu} \epsilon_R^\dagger \eta_R \theta^a +
  \epsilon_R^\dagger \Gamma_\mu{}^\nu \eta_R \dot A_{a\nu} 
  \theta^a~.
\end{equation}
The term $\epsilon_R^\dagger \Gamma_\mu{}^\nu \eta_R$ is a linear
combination of the almost complex structures $I_\mu{}^\nu$, $J_\mu{}^\nu$
and $K_\mu{}^\nu$:
\begin{equation}
  \epsilon_R^\dagger \Gamma_\mu{}^\nu \eta_R = \varepsilon_1
  I_\mu{}^\nu + \varepsilon_2 J_\mu{}^\nu + \varepsilon_3 K_\mu{}^\nu~,
\end{equation}
whence
\begin{equation}
  \dot A_{a\mu} \delta_\epsilon X^a  = \left(\varepsilon^1
    I_\mu{}^\nu + \varepsilon^2 J_\mu{}^\nu + \varepsilon^3
    K_\mu{}^\nu \right) \dot A_{a\nu} \theta^b
 + \epsilon_R^\dagger \eta_R \dot A_{a\mu} \theta^a~.
\end{equation}
From equation \eqref{eq:endoontan}, we may write the action of these
complex structures on $\dot A_{a\nu}$ in terms of the almost complex
structures $\eI$, $\eJ$, $\eK$ on $T_\CC\eM$.  The end result is that
\begin{equation}
  \label{eq:prelinsusy}
  \dot A_{a\mu} \delta_\epsilon X^a = \left(\varepsilon^1 \eI_b{}^a
  + \varepsilon^2 \eJ_b{}^a + \varepsilon^3 \eK_b{}^a
  + \varepsilon^4 \II_b{}^a \right) \dot A_{a\mu} \theta^b~,
\end{equation}
where we have defined $\varepsilon^4 = \epsilon_R^\dagger \eta_R$.  We
remark that the $\varepsilon^{1,2,3,4}$ are  Grassmann odd since so is
$\epsilon_R$.  Since the $\dot A_{a\mu}$ are linearly independent,
equation~\eqref{eq:prelinsusy} is equivalent to
\begin{equation}
  \label{eq:linbossus}
  \delta_\epsilon X^a = \left(\varepsilon^1 \eI_b{}^a
  + \varepsilon^2 \eJ_b{}^a + \varepsilon^3 \eK_b{}^a
  + \varepsilon^4 \II_b{}^a \right) \theta^b~,
\end{equation}
which defines the supersymmetry transformations for the bosonic moduli
$X^a$.

It should be possible to derive the supersymmetry transformations for
the fermionic moduli $\theta^a$ from the gauge theory as well, but we
have been unable to do this and instead we will derive them by
demanding the closure of the supersymmetry algebra.

\subsection{Closure of the moduli space supersymmetry algebra}
\label{sec:closure}

We shall now constrain the geometry of the moduli space by demanding
closure of the supersymmetry algebra.  In contrast with the case of
euclidean monopoles, where the geometry of the moduli is constrained by
demanding the invariance under supersymmetry of the effective action
for the zero modes, the lack of convergence of the $L^2$ metric means
that we cannot write down an action for the zero modes.  It is the
closure of the supersymmetry on the zero modes which will give us 
geometrical information.

To this end let us define odd derivations $\delta_A$, $A = 1,\dots,4$,
by $\delta_\epsilon X^a = \varepsilon^A \delta_A X^a$; that is,
\begin{equation}
  \label{eq:susytranbos}
  \delta_A X^a = \theta^b \eE_A{}_b{}^a~,
\end{equation}
where $\eE_A = (\eI, \eJ, \eK, \II)$, or completely explicitly,
\begin{equation}
  \delta_1 X^a = \theta^b \eI_b{}^a \qquad 
  \delta_2 X^a = \theta^b \eJ_b{}^a \qquad 
  \delta_3 X^a = \theta^b \eK_b{}^a \qquad 
  \delta_4 X^a = \theta^a~.
\end{equation}
Hyperbolic monopoles are half-BPS, whence they preserve 4 of the 8
supercharges of the supersymmetric Yang--Mills theory and this means
that the supersymmetry on the zero modes should close on the
one-dimensional $N=4$ supersymmetry algebra:
\begin{equation}
  \label{eq:susyalg}
  \delta_A \delta_B + \delta_B \delta_A = 2 i \delta_{AB}
  \frac{d\hphantom{t}}{dt}~,
\end{equation}
where $t$ parametrises the curves $X(t),\theta(t)$.  We shall denote
the action of $\frac{d\hphantom{t}}{dt}$ by a prime.

Imposing this on $X^a$ will determine the supersymmetry
transformations of the fermionic moduli $\theta^a$.  For example,
\begin{equation}
  \delta_4^2 X^a = i X'^a \implies \delta_4 \theta^a = i X'^a~,
\end{equation}
and also
\begin{equation}
  \delta_1^2 X^a = i X'^a \implies \delta_1 \theta^a = -i X'^b
  \eI_b{}^a - \theta^b\theta^d \partial_c\eI_b{}^e \eI_d{}^c \eI_e{}^a~,
\end{equation}
and similarly for $\delta_2$ and $\delta_3$ by replacing $\eI$ by
$\eJ$ and $\eK$, respectively.  Next we impose $\delta_4 \delta_i X^a
= - \delta_i \delta_4 X^a$ for $i=1,2,3$.  For example,
\begin{equation}
  0 = \delta_1 \delta_4 X^a + \delta_4 \delta_1 X^a = \theta^d
  \theta^b \left(\partial_d\eI_b{}^a + \partial_c \eI_b{}^e \eI_d{}^c \eI_e{}^a\right)~,
\end{equation}
and similarly for $\eJ$ and $\eK$.  This allows to rewrite in a
slightly simpler way the supersymmetry transformations for the
$\theta^a$:
\begin{equation}
  \label{eq:susytransfer}
  \begin{aligned}[m]
    \delta_1 \theta^a &= -i X'^b \eI_b{}^a + \theta^b \theta^c \partial_c \eI_b{}^a\\
    \delta_2 \theta^a &= -i X'^b \eJ_b{}^a + \theta^b \theta^c \partial_c \eJ_b{}^a\\
    \delta_3 \theta^a &= -i X'^b \eK_b{}^a + \theta^b \theta^c \partial_c \eK_b{}^a\\
    \delta_4 \theta^a &= i X'^a~,
  \end{aligned}
\end{equation}
together with the conditions
\begin{equation}
  \begin{aligned}[m]
    \partial_{[b} \eI_{c]}{}^a - \partial_d \eI_{[b}{}^e \eI_{c]}{}^d \eI_e{}^a &= 0\\
    \partial_{[b} \eJ_{c]}{}^a - \partial_d \eJ_{[b}{}^e \eJ_{c]}{}^d \eJ_e{}^a &= 0\\
    \partial_{[b} \eK_{c]}{}^a - \partial_d \eK_{[b}{}^e \eK_{c]}{}^d \eK_e{}^a &= 0~.
  \end{aligned}
\end{equation}
Multiplying each equation by the corresponding almost complex structure
$\eE_a{}^f$, we obtain the equivalent conditions
\begin{equation}
  \begin{aligned}[m]
    \partial_{[b} \eI_{c]}{}^a \eI_a{}^f + \partial_d \eI_{[b}{}^f \eI_{c]}{}^d &= 0\\
    \partial_{[b} \eJ_{c]}{}^a \eJ_a{}^f + \partial_d \eJ_{[b}{}^f \eJ_{c]}{}^d &= 0\\
    \partial_{[b} \eK_{c]}{}^a \eK_a{}^f + \partial_d \eK_{[b}{}^f \eK_{c]}{}^d &= 0\\
  \end{aligned}
\end{equation}
Comparing with equation~\eqref{eq:Nbracket} in
Appendix~\ref{sec:frol-nijenh-brack}, we see that these conditions are
precisely the vanishing of the following Frölicher--Nijenhuis brackets
$[\eI,\eI]=0$, $[\eJ,\eJ]=0$ and $[\eK,\eK]=0$, which are precisely
the vanishing of the Nijenhuis tensors of the corresponding almost
complex structures.  In other words, $\eI$, $\eJ$ and $\eK$ are
(integrable) complex structures.

Finally we consider the relations imposed by $\delta_i\delta_j X^a = -
\delta_j\delta_i X^a$, for $i,j=1,2,3$ but $i\neq j$.  For example,
\begin{equation}
  \begin{split}
    0 &= \delta_1 \delta_2 X^a + \delta_2\delta_1 X^a\\
    &= \theta^f\theta^d \left(\eI_d{}^c \partial_c \eJ_f{}^a +
    \eJ_d{}^c \partial_c \eI_f{}^a - \partial_d \eI_f{}^b \eJ_b{}^a
    - \partial_d \eJ_f{}^b \eI_b{}^a \right)~,
  \end{split}
\end{equation}
and similarly for the two pairs $(i,j)=(2,3),(3,1)$.  Comparing with
equation~\eqref{eq:FNbracketCoords} in Appendix~\ref{sec:frol-nijenh-brack}, we see that these
conditions are precisely the vanishing of the following Frölicher--Nijenhuis
brackets $[\eI,\eJ]=0$, $[\eJ,\eK]=0$ and $[\eK,\eI]=0$.

Closure of the algebra on the fermionic moduli imposes no further
constraints on the geometry, as we now show.  First we consider
\begin{equation}
  \delta_4 \delta_4 \theta^a = \delta_4 \delta_4 \delta_4 X^a =
  \delta_4 (i X'^a) = i \theta'^a~,
\end{equation}
where we have used that $\delta_4$ and $\frac{d}{dt}$ commute on $X^a$ and,
being derivations, on any differentiable function of $X^a$.  In
particular this implies that $\delta_4^2 = i \frac{d}{dt}$ on any
(differentiable) function of $X$ and $\theta$.  Now let
us consider, for example,
\begin{equation}
  \delta_1\delta_4 \theta^a = \delta_1 \delta^2_4 X^a = \delta_1 (i
  X'^a) = i (\delta_1 X^a)'~,
\end{equation}
whereas on the other hand
\begin{equation}
  \delta_4 \delta_1 \theta^a = \delta_4 \delta_1 \delta_4 X^a =
  - \delta_4^2 \delta_1 X^a = - i (\delta_1 X^a)~,
\end{equation}
where we have used that $\delta_4^2 = i\frac{d}{dt}$ on $\delta_1 X^a$.
Therefore we see that $\delta_1 \delta_4 \theta^a + \delta_4
\delta_1\theta^a = 0$ and similarly for $\delta_2$ and $\delta_3$.
This means that for all $i=1,2,3$, $\delta_i\delta_4 + \delta_4
\delta_i = 0$ on any (differentiable) function of $X^a$ and
$\theta^a$.  Now consider
\begin{equation}
  \delta_1^2 \theta^a = \delta_1^2 \delta_4 X^a = -\delta_1 \delta_4
  \delta_1 X^a = + \delta_4 \delta_1^2 X^a = i \delta_4 X'^a = i
  \theta'^a~.
\end{equation}
Similar calculations show that $\delta_i^2 \theta^a = i \theta'^a$ for
$i=1,2,3$, whence $\delta_i^2 = i \frac{d}{dt}$ on any (differentiable)
function of $X^a$ and $\theta^a$.  Finally, consider
\begin{multline}
  \delta_1 \delta_2 \theta^a = \delta_1 \delta_2 \delta_4 X^a = -
  \delta_1 \delta_4 \delta_2 X^a = + \delta_4 \delta_1 \delta_2 X^a\\ =
  - \delta_4 \delta_2 \delta_1 X^a = + \delta_2 \delta_4 \delta_1 X^a
    = - \delta_2 \delta_1 \delta_4 X^a = - \delta_2 \delta_1 \theta^a~,
\end{multline}
and similarly for the other combinations, whence we see that $\delta_i
\delta_j \theta^a = - \delta_j \delta_i \theta^a$ for $i\neq j$.

In summary, if $\eE$ is any linear combination $\eE = \alpha \eI +
\beta \eJ + \gamma \eK$, then $[\eE,\eE]=0$ and if in addition, $\alpha^2
+ \beta^2 + \gamma^2 = 1$, so that $\eE$ is an almost complex
  structure, the condition $[\eE, \eE]=0$ says that it is integrable.
  Hence the complexified tangent bundle to the hyperbolic monopole
  moduli space has a 2-sphere worth of integrable complex structures
  which act complex linearly. In other words, $\eM$
  has a \emph{pluricomplex structure}, a concept introduced in
  \cite{Bielawski2013}, and which we have hereby shown to follow
  naturally from supersymmetry.

\section*{Acknowledgments}

It is a pleasure to thank Michael Atiyah, Roger Bielawski and Michael
Singer for very stimulating conversations.  A preliminary version of
the work described here was presented in October 2012 by the first
author at the ``Cycles, calibrations and nonlinear partial
differential equations'' conference celebrating the 70th birthday of
Blaine Lawson at the Simons Center in Stony Brook, New York, and he
would like to thank the organisers, particularly Michael Anderson, for
the invitation to speak.  This work was supported in part by the grant
ST/J000329/1 ``Particle Theory at the Tait Institute'' from the UK
Science and Technology Facilities Council.  Finally, we are grateful
to an anonymous referee for comments and suggestions which we believe
have improved the paper.

\appendix

\section{The Frölicher--Nijenhuis bracket of endomorphisms}
\label{sec:frol-nijenh-brack}

The Frölicher--Nijenhuis bracket defines graded Lie superalgebra
structure on the space $\Omega^\bullet(M;TM)$ of vector-valued
differential forms on a manifold $M$.  For a modern treatment see
\cite[Chapter 8]{MR1202431}.  This bracket extends the Lie bracket of
vector fields, thought of as elements of $\Omega^0(M;TM)$.
Endomorphisms of $TM$ can be thought of as elements of
$\Omega^1(M;TM)$ and the Frölicher--Nijenhuis bracket defines a
symmetric bilinear map $[-,-]: \Omega^1(M;TM)\times \Omega^1(M;TM) \to
\Omega^2(M;TM)$.  Paragraph 8.12 in \cite{MR1202431} gives an explicit
expression of the Frölicher--Nijenhuis bracket $[K,L]$ of two
endomorphisms $K,L$ in terms of the Lie bracket of vector fields:
namely,
\begin{multline}
  \label{eq:FNbracket}
  [K,L](X,Y) = [KX,LY] - [KY,LX] - L[KX,Y] + L[KY,X]\\
  - K[LX,Y] + K[LY,X] + (LK+KL)[X,Y]~.
\end{multline}
Applying this to $X = \partial_a$ and $Y
= \partial_b$, we find
\begin{equation}
  \label{eq:FNbracketCoords}
  \begin{split}
    [K,L](\partial_a,\partial_b) &=
    [K_a{}^c\partial_c,L_b{}^d\partial_d] -
    [K_b{}^c\partial_c,L_a{}^d\partial_d] -
    L[K_a{}^c\partial_c,\partial_b]\\
    & \quad {} + L[K_b{}^c\partial_c,\partial_a] -
    K[L_a{}^c\partial_c,\partial_b] +
    K[L_b{}^c\partial_c,\partial_a]\\
    &= \left( K_a{}^c\partial_cL_b{}^d - L_b{}^c\partial_c K_a{}^d -
      K_b{}^c\partial_cL_a{}^d + L_a{}^c\partial_c K_b{}^d \right. \\
      & \quad \left. {} + \partial_b K_a{}^c L_c{}^d - \partial_a K_b{}^c L_c{}^d
      + \partial_b L_a{}^c K_c{}^d - \partial_a L_b{}^c K_c{}^d \right) \partial_d~.
  \end{split}
\end{equation}

It is perhaps easier to remember the case $K=L$:
\begin{equation}
  \label{eq:FNbracketPol}
  \tfrac12 [K,K](X,Y) = [KX,KY] - K[KX,Y] + K[KY,X] + K^2[X,Y]~,
\end{equation}
from which we can recover the general case by the standard
polarisation trick.  Applying this to $X = \partial_a$ and $Y
= \partial_b$, we find
\begin{equation}
  \label{eq:Nbracket}
  \begin{split}
    \tfrac12 [K,K](\partial_a,\partial_b) &=
    [K_a{}^c\partial_c,K_b{}^d\partial_d] -
    K[K_a{}^c\partial_c,\partial_b] +
    K[K_b{}^c\partial_c,\partial_a]\\
    &= \left(K_a{}^c \partial_c K_b{}^d - K_b{}^c \partial_c K_a{}^d
    - \partial_b K_a{}^c K_c{}^d +  \partial_a K_b{}^c K_c{}^d \right) \partial_d~.
  \end{split}
\end{equation}


\bibliographystyle{utphys}
\bibliography{SusyHypMon,Duality}

\end{document}